%% file: magnetic_rev_rev.tex
\documentclass[graybox]{svmult}

\usepackage{mathptmx}       
\usepackage{helvet}         
\usepackage{courier}        
\usepackage{type1cm}        
\usepackage{makeidx}         
\usepackage{graphicx}        
\usepackage{multicol}        
\usepackage[bottom]{footmisc}

\newcommand{\pbp}{\langle \bar \psi \psi \rangle}
\newcommand{\beq}{\begin{eqnarray}}
\newcommand{\eeq}{\end{eqnarray}}

\newcommand{\real}{{\sf I}\kern-.12em{\sf R}}
\newcommand{\comp}{{\sf I}\kern-.50em{\sf C}}
\newcommand{\unity}{{\sf I}\kern-.54em{\sf 1}}

\newcommand{\tef}{\theta_{\rm eff}}

\newcommand{\esb}{\vec E \cdot \vec B}

\newcommand{\ceb}{\chi_{CP}\,}

\makeindex             

\begin{document}

\title*{Lattice QCD Simulations in External Background Fields}
\author{Massimo D'Elia}
\institute{Massimo D'Elia \at Dipartimento di Fisica dell'Universit\`a
di Pisa and INFN - Sezione di Pisa,\\
 Largo Pontecorvo 3, I-56127 Pisa, Italy\\
\email{delia@df.unipi.it}}
%
%
\maketitle

\abstract*{
We discuss recent results and future prospects 
regarding the investigation, by lattice simulations, of the non-perturbative properties
of QCD and of its phase diagram in presence of magnetic
or chromomagnetic background fields. After a brief introduction to
the formulation of lattice QCD in presence of external fields, we focus 
on studies regarding the effects of external fields on chiral symmetry breaking,
on its restoration at finite temperature and on deconfinement. We conclude with
a few comments regarding the effects of electromagnetic background fields on
gluodynamics.}

\abstract{
We discuss recent results and future prospects 
regarding the investigation, by lattice simulations, of the non-perturbative properties
of QCD and of its phase diagram in presence of magnetic
or chromomagnetic background fields. After a brief introduction to
the formulation of lattice QCD in presence of external fields, we focus 
on studies regarding the effects of external fields on chiral symmetry breaking,
on its restoration at finite temperature and on deconfinement. We conclude with
a few comments regarding the effects of electromagnetic background fields on
gluodynamics.}

\section{Introduction}
\label{sec:1}

The properties of strong interactions in presence of strong external fields
are of great phenomenological and theoretical importance.
Large magnetic or 
chromomagnetic background fields, of the order of 
$10^{16}$ Tesla, i.e. $\sqrt{|e|B} \sim 1.5$ GeV, may have been 
produced at the time of the cosmological electroweak phase 
transition~\cite{Vachaspati:1991nm} and may
have influenced the subsequent evolution of the Universe,
in particular the transition from deconfined to confined strongly
interacting matter.
Slightly lower magnetic fields 
are expected to be produced in non-central heavy
ion collisions (up to $\sim  10^{15}$
Tesla at LHC~\cite{cme0,heavyionfield2}), where they may give 
rise, in presence of non-trivial topological vacuum fluctuations, 
to CP-odd effects consisting in the separation of electric charge
along the direction of the background magnetic field, the 
so-called {\em chiral magnetic effect}~\cite{vilenkin,cme0,cme1}.
Finally, large magnetic (or even chromomagnetic~\cite{pstar}) 
fields
are expected in some compact
astrophysical objects, such as  
magnetars~\cite{magnetars} (see \cite{mereghetti} for a recent 
review on the subject).

Apart from the phenomenological issues above, there is 
great interest also at a purely theoretical level,
since background fields can be useful probes to ge insight
into  the properties of strong interactions and the 
non-perturbative structure of the QCD vacuum.
A typical example is chiral symmetry breaking, which 
is predicted to be enhanced by the presence of a magnetic
background, a phenomenon which is known as magnetic 
catalysis~\cite{catal_1,catal_2,catal_3,catal_4,catal_5,catal_6,catal_7,catal_8,catal_9,catal_10,
smilga,cohen,
catal1_1,catal1_2,catal1_3,catal1_4,catal1_5,catal1_6,catal1_a,catal1_b,catal1_7,catal1_8,catal1_9,catal1_10,
catal1_12,catal1_c,shov,anisotropic,schmitt}.

Such interest justifies the large efforts which have been dedicated
in the recent past to this subject by a variety
of different approaches, most of which are reviewed in the present 
volume.
Lattice QCD represents, in general, the ideal tool for a first principle 
investigation of the non-perturbative properties of strong interactions.
This is particularly true for the study of QCD in presence of a strong
external magnetic or chromomagnetic field, since no kind of technical
problem, such as a sign problem, appears, preventing
standard Monte-Carlo simulations, as it happens instead
at finite baryon chemical potential.

Similarly to the continuum theory, 
the magnetic field is introduced on the lattice
in terms of additional $U(1)$ degrees of freedom 
(see Sec.~\ref{subsec:2}), which are not directly
coupled to the original $SU(3)$ link variables and  
affect quark propagation by a modification of the 
covariant derivative (i.e. the fermion matrix).
The fermion determinant, contrary to
the case of a baryon chemical potential $\mu_B$ or of an electric 
background field, is still real and positive, allowing for a 
probabilistic interpretation of the path integral measure.
Lattice studies including the presence of electromagnetic fields 
have been done since long, originally with the purpose
of studying the electromagnetic properties of 
hadrons~\cite{hadron1,hadron2,detmold,tiburzi,alexandru}. 
The introduction of a chromomagnetic field requires
a different approach, since in this case the background variables
are strictly related to the quantum gluon degrees of freedom:  
a standard procedure is that defined in 
the
framework of the lattice Schr\"odinger 
functional~\cite{sch_1,sch_2,cea_old,cea_polosa,
cea_quench_1,cea_quench_2,cea_full,antropov,cea_monofull}
(see Sec.~\ref{subsec:3}).

The recent interest about external field effects in the QCD vacuum and
around the deconfining transition has stimulated lattice investigations
by many 
groups~\cite{blum,itep1,itep11,itep3,itep4,demusa,itep2,dene,itep5,yamamoto,bali,
ilgenfritz,luschev,bali2,bali3}.
Several studies have followed a quenched or partially quenched approach, 
considering only the effect of the magnetic background on physical observables,
i.e. on quarks propagating in the background of the given magnetic field
and of non-Abelian gauge configurations, the latter being sampled 
without taking into account the magnetic field.

However, other studies 
have shown that the electromagnetic background can have, via
quark loop effects, a strong influence also on the distribution
of non-Abelian fields, thus requiring
an unquenched approach. A considerable
contribution to magnetic catalysis appears to be related to 
gluon field modifications~\cite{dene}; the magnetic field clearly
affects also gluonic observables, like the plaquette~\cite{ilgenfritz}
or those typically
related to confinement/deconfinement, like the Polyakov 
loop~\cite{demusa,bali,ilgenfritz}.
That claims for a more systematic investigation
about the effects of electromagnetic backgrounds
on the gluonic sector of QCD and about the possible interplay between 
magnetic and chromomagnetic backgrounds, which could have many
implications both at a theoretical and at a phenomenological level
(e.g. for cosmology or heavy ion collisions).

The review is organized as follows. In Sec.~\ref{sec:2} we discuss
the formulation of QCD in external fields, considering both the case
of an electromagnetic and of a chromomagnetic field. 
In Sec.~\ref{sec:vacuum} we review studies regarding vacuum properties
in external fields, in particular concerning magnetic catalysis.
Sec.~\ref{sec:phasediagram} is devoted to a discussion about the 
influence of magnetic or chromomagnetic background fields
on deconfinement and chiral symmetry restoration. In Sec.~\ref{gluodyn}
we extend the discussion
about the influence that electromagnetic fields may have
on the gluonic sector, with a particular emphasys on symmetries and
on suggestions for future studies on the subject.

\section{Background Fields on the Lattice}
\label{sec:2}

In this Section we will briefly review the methods which are commonly adopted 
for a numerical study of QCD in presence of background fields. There
is a difference between the case of electromagnetic and 
chromomagnetic backgrounds: in the former, the external field
simply adds new $U(1)$ degrees of freedom which contribute to quark
propagation but are decoupled from the $SU(3)$ degrees of freedom
and from their updating during the Monte-Carlo simulation; in the latter,
the background field is made up of the same degrees of freedom which are 
dynamically updated during the simulation,
i.e. the $SU(3)$ gauge variables, and one 
has to follow a Schr\"odinger functional approach.

\subsection{Electromagnetic Fields}
\label{subsec:2}

A quark field $q$ propagating in the background of a non-abelian $SU(3)$ gauge
field plus an electromagnetic field is described, in the Euclidean
space-time, by a Lagrangean\
$\bar \psi ( D\hspace{-6pt}\slash  + m) \psi$, where the covariant
derivative 
\beq
D_\mu = \partial_\mu + \I\, g A^a_\mu T^a + \I\, q a_\mu
\eeq
contains contributions from the non-Abelian gauge field 
$A_\mu = A^a_\mu T^a$
and from the Abelian 
gauge field $a_\mu$. Here $T^a$ are the $SU(3)$ generators,
$g$ is the $SU(3)$ gauge coupling and $q$ is the quark electric charge.

Going to a discrete lattice formulation, $SU(3)$ 
gauge invariance is naturally preserved~\cite{wilson} 
by requiring that quarks pick up an appropriate 
non-Abelian phase when hopping
from one site of the lattice to the other, so 
that the gauge field $A_\mu$ is substituted by 
the elementary parallel transports 
$U_\mu (n)$, which reduce to $U_\mu (n) \simeq 1 + \I\, g a A_\mu(n)$
as the lattice spacing $a$ vanishes in the continuum limit.
 
A gauge invariant quantity involving an antiquark at site 
$n$ and a quark at site $n + \hat\mu$ is therefore 
$\bar\psi (n) U_\mu(n) \psi(n + \hat\mu)$
($\hat\mu$ is a unit vector on the lattice).
The same approach can be taken for the Abelian field, so that 
quarks going from site $n$ to site $n + \mu$ pick up also 
the Abelian phase $u_\mu(n) = \exp(\I\, q \int_{an}^{a(n + \hat\mu)} 
d x_\mu a_\mu) \simeq \exp(\I\, e\, q\, a_\mu(n))$.
A possible discretization of the covariant derivative is then
\beq
D_\mu \psi \to  \frac{1}{2a} 
\left(U_\mu(n) u_\mu(n) \psi (n + \hat\mu) - 
U^\dagger_\mu(n-\hat\mu) u^*_\mu(n-\hat\mu) \psi (n - \hat\mu) \right)
\label{covder}
\eeq
Of course, in presence of different quark species, carrying different electric
charges $q$, the $U(1)$ part of the covariant
derivative, $u_\mu(n)$, will change
from quark to quark. Therefore, as usual, the discretized version of
the fermion action is  
a bilinear
form in the fermion fields, $\bar\psi(i) M_{i,j} \psi(j)$, however
the elements of $M$ now belong to $U(3)$ instead of $SU(3)$.

\subsubsection{Magnetic Fields}

To take an explicit example, which is directly linked to most of the 
results that we discuss in the following, let us consider the finite
temperature partition
function  of two degenerate quarks, e.g. the $u$ and $d$ quarks, in presence
of a constant and uniform magnetic background field. We shall consider
a rooted staggered fermion
discretized version of the theory, in which each quark is described
by the fourth root of the determinant of a staggered fermion matrix:
\beq
Z(T,B) \equiv \int \mathcal{D}U e^{-S_{G}} 
\det M^{1\over 4} [B,q_u]
\det M^{1\over 4} [B,q_d]
\:
\label{partfun1}
\eeq
where in the standard formulation
\begin{eqnarray}
M_{i,j} [B,q] = a m \delta_{i,j} 
&+& {1 \over 2} \sum_{\nu=1}^{4} \eta_\nu(i) \left(
u_\nu(B,q)(i)\ U_{\nu}(i) \delta_{i,j-\hat\nu}
\right. \nonumber \\ &-& \left.
u^*_\nu(B,q){(i - \hat\nu)}\ U^{\dag}_\nu{(i-\hat\nu)} \delta_{i,j+\hat\nu} 
\right) \:.
\label{fmatrix1}
\end{eqnarray}
$\mathcal{D}U$ is the functional integration over the non-Abelian gauge link
variables, $S_G$ is the discretized pure gauge action 
 and $u(B,q)_{i,\nu}$ are the 
Abelian links corresponding to the background field;  
$i$ and $j$ refer to lattice sites and
$\eta_{i,\nu}$ are the staggered
phases. 
We shall consider two different charges for the two flavors, 
$q_u = 2|e|/3$ and $q_d = -|e|/3$.
Periodic or antiperiodic boundary conditions along the
Euclidean time direction must be taken, respectively
for gauge or fermion fields, in order to define a thermal theory:
the temperature $T$ corresponds to the inverse temporal extension.

A special discussion must be devoted to the issue of spatial boundary 
conditions. It is well known that, for a given lattice size (which is 
usually constrained by the computational power available) periodic boundary 
conditions in space (for all fields) are the one which best approximate
the infinite volume limit, and are therefore the standard choice in 
lattice simulations. On the other hand, they place some constraints 
on the possible magnetic fields~\cite{bound1,bound2,bound3,wiese}.

Let us consider a magnetic field directed along the $\hat z$ axis
of a three-dimensional torus, $\vec B = B \hat z$,
and let $l_x$ and $l_y$ be the torus
extensions in the orthogonal directions.
The circulation
of $a_\mu$ along any closed path 
in the $x-y$ plane, enclosing a 
region of area $A$, is given by Stokes theorem
\beq
\oint a_\mu d x_\mu = A B \, 
\eeq
On the other hand it is ambiguous, on a closed surface like a torus, 
to state which is the
region enclosed by a given path: we can choose 
the complementary region of area $l_x l_y - A$ and
equally state
\beq
\oint a_\mu d x_\mu = (A - l_x l_y) B \, .
\eeq
The ambiguity can be resolved either by admitting discontinuities 
for $a_\mu$ somewhere on the torus, or by covering the torus with various
patches corresponding to different gauge choices. However it is 
essential to guarantee that 
the ambiguity be not visible by any charged particle moving on the 
torus, i.e. the phase factor picked up by a particle of charge 
$q$ moving along the given path must be a well defined quantity.
Such requirement leads to magnetic field quantization, indeed
it is satisfied only if
\beq
\exp \left( \I\, q B A \right) = \exp \left( \I\, q B (A - l_x l_y) \right) 
\eeq
i.e. if 
\beq
q B= \frac{2\pi b}{l_x l_y}
\label{defb}
\eeq
where $b$ is an integer. It is interesting to notice that this is 
exactly the same argument leading to Dirac quantization of the 
magnetic monopole charge (in that case the closed surface is any
sphere centered around the monopole).
The quantization rule depends on the 
electric charges of the particles moving on the torus~\footnote{
Of course we assume all particles living on the torus 
to carry integer multiples
of some elementary electric charge, otherwise a consistent
quantization of the magnetic field would not be possible.},
in our case it is set by the $d$ quark, which brings the 
smallest charge unit
$q_d = -|e|/3$, hence
\beq
|e| B = \frac{6 \pi b}{l_x l_y} 
= \frac{6 \pi b}{ a^{2} L_x L_y} 
\label{Bquant}
\eeq
where $L_x$ and $L_y$ are the dimensionless lattice 
extensions in the $x,\ y$ directions.

A possible choice for the gauge links, corresponding 
to the continuum gauge field
\beq
a_y = B x
\:; \qquad\qquad
a_\mu = 0
\quad \mathrm{for} \quad \mu = x,z,t
\:.
\label{afield}
\eeq
is the following:
\beq
u_y(B,q)(n) = e^{\I\, a^2 q B\, n_x}
\, ;\ \ \ \ \ \ \
u_\mu(B,q)(n) = 1 
\ \ \  \mathrm{for} \ \ \ \mu = x,z,t\ \, .
\label{u1field}
\eeq
The
smoothness of the background field across the boundaries and 
the gauge invariance of
the fermion action are guaranteed if appropriate boundary conditions
are taken for the fermion fields along the $x$ direction~\cite{wiese},
that corresponds to modifying 
the $U(1)$ gauge links in the $x$ direction as follows:
\beq
u_x(B,q)(n)|_{n_x = L_x} = e^{-\I\, a^2 q L_x B\, n_y} \, .
\label{boundary}
\:
\eeq

A different condition on the possible magnetic field values 
explorable on the lattice is placed by the 
ultraviolet (UV) cutoff, i.e. by discretization itself.
All the information about the presence of the magnetic field is contained
in the phase factors picked up by particles moving on the lattice. There
is a minimum such path on a cubic lattice, the plaquette;
in particular, a particle
moving around a $x-y$ plaquette takes a phase factor 
\beq
\exp (\I\, q a^2 B) = \exp \left( \I\, \frac{2 \pi b}{L_x L_y} \right) \, 
\label{phasefactor}
\eeq
and all other possible phase factors, corresponding to different paths, 
are positive integer powers of that. It is evident that the phase factor in 
Eq.~(\ref{phasefactor})
cannot distinguish magnetic fields such that $ q a^2 B$ differs
by multiples of $2 \pi$: one would need smaller paths to do that, which 
are unavailable because of the UV cutoff. Hence it is possible to define
a sort of first Brillouin zone
for the magnetic field
\beq
- \frac{\pi}{a^2} < q B < \frac{\pi}{a^2} 
\label{Bcutoff}
\eeq
and we expect that physical quantities must 
be periodic in $q B$ with a period $2 \pi /a^2$; such periodicity
is unphysical and one should always be cautious when exploring 
magnetic field values which get close to the limiting values in 
Eq.~(\ref{Bcutoff}).

Another issue which is relevant to the lattice implementation 
of background fields regards 
translational invariance. The magnetic field
$B$ breaks the 
translational invariance of the continuum torus in the $x-y$ directions 
explicitly~\cite{wiese}. That is clearly seen 
by looking at the $U(1)$ Wilson lines (holonomies) $W_x$ and $W_y$, 
i.e. the $U(1)$ parallel transports along straight paths in the
$x$ or $y$ direction and closed around the torus by 
periodic boundary conditions, which are equal to 
\beq
W_x = \exp(-\I\, q B\, l_x\, y)\ \ ; \ \ \ \ \ \ 
W_y = \exp(\I\, q B\, l_y\, x)\ \  \ \ \ \ \ 
\eeq
as can be verified explicitly by means of the elementary parallel
transports given in Eq.~(\ref{u1field}) and (\ref{boundary}). 
The Wilson lines therefore leave only 
a residual simmetry corresponding to discrete translations by multiples
of 
\beq
\tilde{a}_x = \frac{ 2 \pi}{q B l_y} = \frac{l_x}{b}\ \ ; \ \ \ \ \ 
\tilde{a}_y = \frac{ 2 \pi}{q B l_x} = \frac{l_y}{b}\ \ ; 
\eeq
note that Wilson lines are gauge invariant quantities, hence the 
breaking of translational invariance is not a matter of gauge choice.

However, on a discrete cubic lattice, translational invariance
is already broken to a discrete residual group, corresponding to 
multiples of the lattice spacing $a$, hence for a lattice theory in 
presence of a constant magnetic field, translational invariance is further
reduced to discrete steps which are multiples of both 
$\tilde{a}_{x/y}$ and $a$. Since 
$a/\tilde{a}_{x/y} = b/L_{x/y}$, lattice translational
symmetry 
is preserved for $b$ multiple of $L_{x/y}$, but is 
strongly reduced or completely lost for different values, despite
the fact that the magnetic background field is uniform. 
\\

We have discussed so far about one possible lattice implementation
of
the fermionic action  
in presence of a magnetic field, Eqs.~(\ref{covder}) 
and (\ref{fmatrix1}), which is based on the simplest symmetric
discretization of the covariant derivative, which 
takes quarks from nearest neighbours sites, rotating them by 
a non-Abelian and an Abelian phase, both corresponding to the 
simplest straight path, i.e. the elementary parallel transport.
Such kind of discretization has been adopted in many lattice studies,
see e.g.~\cite{demusa,dene,ilgenfritz}.

It is well known, however, that different, improved discretizations
can lead to an improved convergence towards the continuum limit.
Improvement can proceed in different ways: one can 
consider less simple discretizations of the derivative, involving
fermion fields which are not nearest neighbours;
one can also consider and average over paths different from the simplest 
straight one, 
so as to 
smear out UV fluctuations: this is idea adopted, for instance,
in~\cite{bali,bali2,bali3}, where stout smearing has been used. 
If one considers
lattice derivatives involving non-nearest neighbours lattice sites,
one must include appropriate composite $U(1)$ parallel transports,
so as to preserve the $U(1)$ gauge invariance of the fermion action.
Which paths, however, and how the chosen paths are related to the paths
considered for the non-Abelian phases, is a question which leaves
much freedom.

Let us consider, for simplicity, a gauge invariant 
bilinear term involving nearest 
neighbour quarks, $\bar\psi (n) U_\mu(n) \psi(n + \hat\mu)$,
which reduces to $\bar\psi (x) (1 + a D_\mu) \psi(x) + O(a^2)$
in the na\"{\i}ve continuum limit. We can choose a more general,
improved
such term, involving different paths, i.e. we can modify 
\beq
\bar\psi (n) U_\mu(n) \psi(n + \hat\mu)\  \to\ 
\bar\psi (n) \left( \sum_{C(n,n + \hat\mu)} 
\alpha_C\ U_C
\right) \psi(n + \hat\mu) 
\eeq
where $C(n,n + \hat\mu)$ are a set of paths connecting $n$ to $n + \hat\mu$
(e.g. for smearing one includes staples),
$U_C$ are the parallel transports taken along those paths
and the coefficients $\alpha_C$ are chosen in order to keep
the na\"{\i}ve continuum limit unchanged, up to $O(a^2)$.
Let us now consider the inclusion of a $U(1)$ external field: a possible
prescription is that, as quarks explore all considered paths $C$, taking the
associated non-Abelian phases, they also take the corresponding $U(1)$
phases along the same paths, i.e.
\beq
\bar\psi (n) U_\mu(n) u_\mu(n) \psi(n + \hat\mu) \to 
\bar\psi (n) \left( \sum_{C(n,n + \hat\mu)} 
\alpha_C\ U_C\ u_C 
\right) \psi(n + \hat\mu) \, .
\eeq
However, different prescriptions can be taken and, in general, 
the sum over $U(1)$ phases can be decoupled from that on $SU(3)$ phases,
without affecting gauge invariance.
We can consider for example
\beq
\bar\psi (n) \left( \sum_{C(n,n + \hat\mu)} 
\alpha_C\ U_C 
\right) 
\left( \sum_{C'(n,n + \hat\mu)} 
\alpha'_{C'}\ u_C'
\right) 
\psi(n + \hat\mu) 
\eeq
where $C'$ is a different set of paths. This is the choice made, for instance,
in~\cite{bali,bali2,bali3}, where stout smearing is applied to $SU(3)$ links,
while the $U(1)$ phases are left unchanged (i.e. $C'$ runs over the
elementary straight path only); that seems a reasonable choice, since 
the $U(1)$ field is already a classical smooth field.

It can be easily shown that for a uniform background and for free fermions, i.e. 
in absence of non-Abelian fields, all possible different choices 
of paths for the $U(1)$ term are equivalent, apart from a global
phase stemming from interference (Aharonov-Bohm like) effects
among the different paths. In presence of non-Abelian fields, however,
this is in general not true and it would be interesting to explore
the systematics connected to different choices.
\\

A final comment concerns the issue of possible renormalizations
related to the introduction of the external magnetic field.
The quantity entering the covariant derivative 
and quark dynamics is the combination $q a_\mu$, which 
does not renormalize. The situation is clearer in the lattice
formulation, where the objects carrying information about
the external field are the parallel transports of the 
$U(1)$ gauge field along closed loops (think of the 
loop expansion of the fermionic determinant). Such 
phase factors are gauge inviariant quantities and 
cannot renormalize. 
The only remaining possibility is that
the lattice spacing $a$ itself depends on the 
magnetic field $B$, because of external field effects
at the scale of the UV cutoff, so that the physical area
enclosed by a given loop changes, leading to an 
effective external field renormalization.
Such possibility, however, has been excluded
by a detailed study presented in~\cite{bali},
showing that $a$ does not depend on $B$ within errors.

\subsubsection{Electric Fields}

After the inclusion of a background magnetic field, the 
$D\hspace{-6pt}\slash$ operator is still anti-Hermitean
and it still anticommutes with $\gamma_5$, hence its
spectrum is purely imaginary with non-zero eigenvalues
coming in conjugate pairs, so that 
$\det (D\hspace{-6pt}\slash + m) > 0$ and Monte-Carlo simulations
are feasible, i.e. the path integral measure can be interpreted as a 
probability distribution function over gauge configurations.

The situation is different in presence of background electric 
fields~\cite{shintani,alexandru,tiburzi}.
Let us consider a constant electric field $\vec E = E \hat z$, there are various 
possible choices for a corresponding vector potential in Minkowski space,
like $(A_t = - E z,\ A_i = 0)$ or  $(A_z = - E t,\ A_x = A_y = A_t = 0)$.
In every case, after continuation to Euclidean space time, 
$t \to -\I\, \tau$ and $A_t \to \I\, A_0$, the vector potential becomes
purely imaginary, thus destroying the anti-Hermitean properties
of the Dirac matrix.

That means that the fermion determinant is complex and numerical simulations
are not feasible. This "sign problem" has a very strict relation with the usual
sign problem which is encountered in the study of QCD at finite
baryon chemical potential, indeed also a baryon chemical potential
can be viewed as a constant background potential $A_0 \neq 0$.
As for finite density QCD, also in this case a possible way out 
is to consider a purely imaginary electric field: the lattice formulation 
is then completely analogous to that of a magnetic field.
For example, a purely imaginary electric field along the $\hat z$ direction
can be formulated on the lattice following 
Eqs.~(\ref{u1field}) and (\ref{boundary}), by just replacing replacing $(x,y) \to (z,t)$ and 
$B \to {\rm Im} (E)$. A similar approach, followed by analytic continuation
to real electric fields,
is that usually adopted for the study of hadron electric polarizabilities.

\subsection{Chromomagnetic Background Fields}
\label{subsec:3}

As we have discussed above, the implementation of an electromagnetic 
background field amounts to adding extra $U(1)$ degrees of freedom to the 
Dirac matrix: the added $U(1)$ field is a classic, static field which is 
decoupled from the $SU(3)$ gauge field appearing in the functional 
integration. The situation is different for a color background
field, since in this case one would like to consider quantum fluctuations
of the non-Abelian field around a given backgroud, e.g. a static
ad uniform chromo-magnetic field, however  
the functional integration over gauge variables 
can destroy information about the background field completely.

A common way to approach the problem 
is based on the Schr\"odinger functional approach~\cite{sch_1,sch_2}:
one considers functional integration over Euclidean space-time
with fixed temporal boundaries, $\tau_1$ and $\tau_2$, and 
over gauge configurations
which are frozen to particular assigned values, 
$A^{ext1}_i(\vec x,\tau_1)$ and
$A^{ext2}_i(\vec x,\tau_2)$, at the initial and final times.
This is related to the quantum amplitude of passing from 
the field eigenstate $|A^{ext1}_i\rangle$ to the field 
eigenstate $|A^{ext2}_i\rangle$ in the time 
$(\tau_2 - \tau_1)$. The amplitude is dominated, in the classical
limit, by the gauge configuration which has the
minimal action among those with the given boundary conditions; 
functional integration can then be viewed as an integration over
quantum fluctuations around this classical background field.

In Refs.~\cite{cea_old,cea_polosa,cea_quench_1,cea_quench_2,cea_full} 
such formalism has been
implemented on the lattice by considering  equal initial and final
fields, $A^{ext1}_i(\vec x,\tau_1) = 
A^{ext2}_i(\vec x,\tau_2) = A^{\rm ext}_i(\vec x)$.
In this way one can define 
a lattice gauge invariant
effective action $\Gamma[\vec{A}^{\rm ext}]$ for the external background
field $\vec{A}^{\rm ext}$:
\beq
\label{Gamma}
\Gamma[\vec{A}^{\rm ext}] = -\frac{1}{L_t} \ln
\left\{
\frac{{\mathcal{Z}}[\vec{A}^{\rm ext}]}{{\mathcal{Z}}[0]}
\right\}
\eeq
where $L_t$ is the lattice temporal extension and
${\mathcal{Z}}[\vec{A}^{\rm ext}]$ is the
lattice functional integral
\begin{eqnarray}
\label{ZetaT}
\mathcal{Z}_T \left[ \vec{A}^{\rm ext} \right]  &=
&\int_{U_k(L_t,\vec{x})=U_k(0,\vec{x})=U^{\rm ext}_k(\vec{x})}
\mathcal{D}U \,  {\mathcal{D}} \psi  \, {\mathcal{D}} \bar{\psi} e^{-(S_W+S_F)}
\nonumber \\&=&  \int_{U_k(L_t,\vec{x})=U_k(0,\vec{x})=U^{\rm ext}_k(\vec{x})}
\mathcal{D}U e^{-S_W} \, \det M \,,
\end{eqnarray}
where $S_G$ and $S_F$ 
are the pure gauge and fermion action respectively, 
$M$ is the fermionic matrix.
The functional integration is performed over the lattice links,
with the constraint
\beq
\label{coldwall}
U_k(\vec{x},x_t=0) = 
U_k(\vec{x},x_t=L_t) = 
U^{\rm ext}_k(\vec{x})
\,,\,\,\,\,\, (k=1,2,3) \,\,,
\eeq
where $U^{\rm ext}_k(\vec{x})$ are the elementary parallel transports
corresponding to the continuum
gauge potential $\vec{A}^{\rm ext}(x)=\vec{A}^{\rm ext}_a(x) \lambda_a/2$;
fermion fields as well as temporal links are left unconstrained.
${\mathcal{Z}}[0]$ is defined as in Eq.~(\ref{ZetaT}), but adopting a zero
external field.

In the following we shall consider the case of an Abelian static
and uniform chromomagnetic background field: in this case 
one can choose $\lambda_a$ belonging to the Cartan subalgebra of the 
gauge group, e.g. $\lambda_a = \lambda_3$, while the 
explicit form of $\vec{A}^{\rm ext}_a(x)$, hence of the lattice
links, can be chosen in the same way as for a standard magnetic field,
see Eqs.~(\ref{afield}), (\ref{u1field}) and (\ref{boundary}).
In this case, as well as in other cases in which 
the static background field does not vanish at spatial infinity, one
usually imposes also that spatial links exiting
from sites belonging to the spatial boundaries  are fixed, at all times,
according to 
Eq.~(\ref{coldwall}): that corresponds to the 
requirement that fluctuations over the background field vanish at infinity.

Notice that a colored background field influences directly,
and through the same gauge coupling $g$, 
the dynamics of both quark and gluon fields. 
Indeed numerical simulations in presence of colored background fields have
considered originally just the case of pure Yang-Mills theories.

Differently from 
the usual formulation of the lattice Schr\"odinger
functional~\cite{sch_1,sch_2}, where a cylindrical
geometry is adopted, the effective action defined 
by Eq.~(\ref{Gamma}) assumes an
hypertoroidal geometry, i.e.
the first and the last time slice are identified and periodic boundary
conditions are assumed in the time direction for gluon fields.
For finite values of $L_t$, having adopted also the prescription
of anti-periodic boundary conditions in time direction for quark fields,
Eq.~(\ref{ZetaT}) can be interpreted as the
thermal partition function
${\mathcal{Z}_{T}}[\vec{A}^{\rm ext}]$
in presence of the background field $\vec{A}^{\rm ext}$, with
the temperature given by $T=1/(a L_t)$.
The  gauge invariant
effective action, Eq.~(\ref{Gamma}), is then replaced by 
the free energy functional:
\beq
\label{freeenergy}
{\mathcal{F}}[\vec{A}^{\rm ext}] = -\frac{1}{L_t} \ln
\left\{
\frac{{\mathcal{Z}_{T}}[\vec{A}^{\rm ext}]}{{\mathcal{Z}_{T}}[0]}
\right\} \; .
\eeq

\section{Vacuum Properties in Backgroud Fieds: Magnetic Catalysis}
\label{sec:vacuum}

One of the most significant effects that a magnetic background field can induce 
on the QCD vacuum, as well as on other systems characterized by the 
chiral properties of fermion fields, 
is known as magnetic catalysis.
It consists in an enhancement of chiral symmetry breaking, or 
spontaneous mass generation, which can be thought as related
to the dimensional reduction taking place in the dynamics of particles moving in a 
strong external magnetic field (see~\cite{shov} for a recent comprehensive review).

The enhancement of chiral symmetry breaking reveals itself in a dependence
of the chiral condensate $\pbp$ on the magnetic field $B$; there
are several predictions for the actual functional form $\pbp(B)$, depending
on the adopted model, however some general features can be summarized as 
follows. By charge conjugation symmetry, 
$\pbp$ must be an even function of $B$, hence, if
the theory is analytic around $B = 0$, one expects that 
$\pbp$ depends quadratically on $B$ in a regime of small enough fields.
The analyticity assumption is violated in presence of charged 
massless fermions,
indeed
chiral perturbation theory~\cite{smilga} predicts a linear behavior
in $B$ if $m_\pi = 0$; such linear dependence is recovered in computations
at $m_\pi \neq 0$, in the limit
$e B \gg m_\pi^2$~\cite{cohen}. Different power behaviors in $B$ 
can be found in other model computations.
In the following we shall make use of 
the relative increment of the chiral
condensate, defined as
\beq
r (B) = \frac{\pbp (B) - \pbp(B = 0)}{\pbp (B = 0)} \, .
\label{rbdef}
\eeq

Early lattice investigations of magnetic catalysis have been done in the 
quenched approximation, both for $SU(2)$~\cite{itep1}
and for $SU(3)$~\cite{itep2} gauge theories. In this case, the chiral condensate
is computed by inverting a Dirac operator which contains the contribution
from the external field, but on gauge configurations
sampled in absence of dynamical fermion contributions.
The outcome is that $r(B) \propto B$ for $SU(2)$~\cite{itep1}, while 
$r(B) \propto B^\nu$, with $\nu \sim 1.6$, for $SU(3)$~\cite{itep2}.

On the other hand,
the inclusion of contributions
from dynamical fermions moving in the background field may produce significant
effects.
The chiral condensate is related, via the 
Banks - Casher relation~\cite{banks},
to the density $\rho(\lambda)$
of eigenvalues of the Dirac operator around $\lambda = 0$: 
$\pbp = \pi \rho(0)$. Such density can change, as a function of $B$, 
both because the Dirac operator definition itself changes, 
and because the distribution
of gauge configurations, over which the operator is computed, 
is modified by dynamical fermion contributions.
Studies of magnetic catalysis involving full dynamical simulations
have been reported in~\cite{dene,bali2}, for QCD with 3 colors,
and in~\cite{ilgenfritz} for QCD with 2 colors.

\begin{figure}[t!]
$\, $
\vspace{1.0cm}
\begin{center}
\includegraphics[scale=.46]{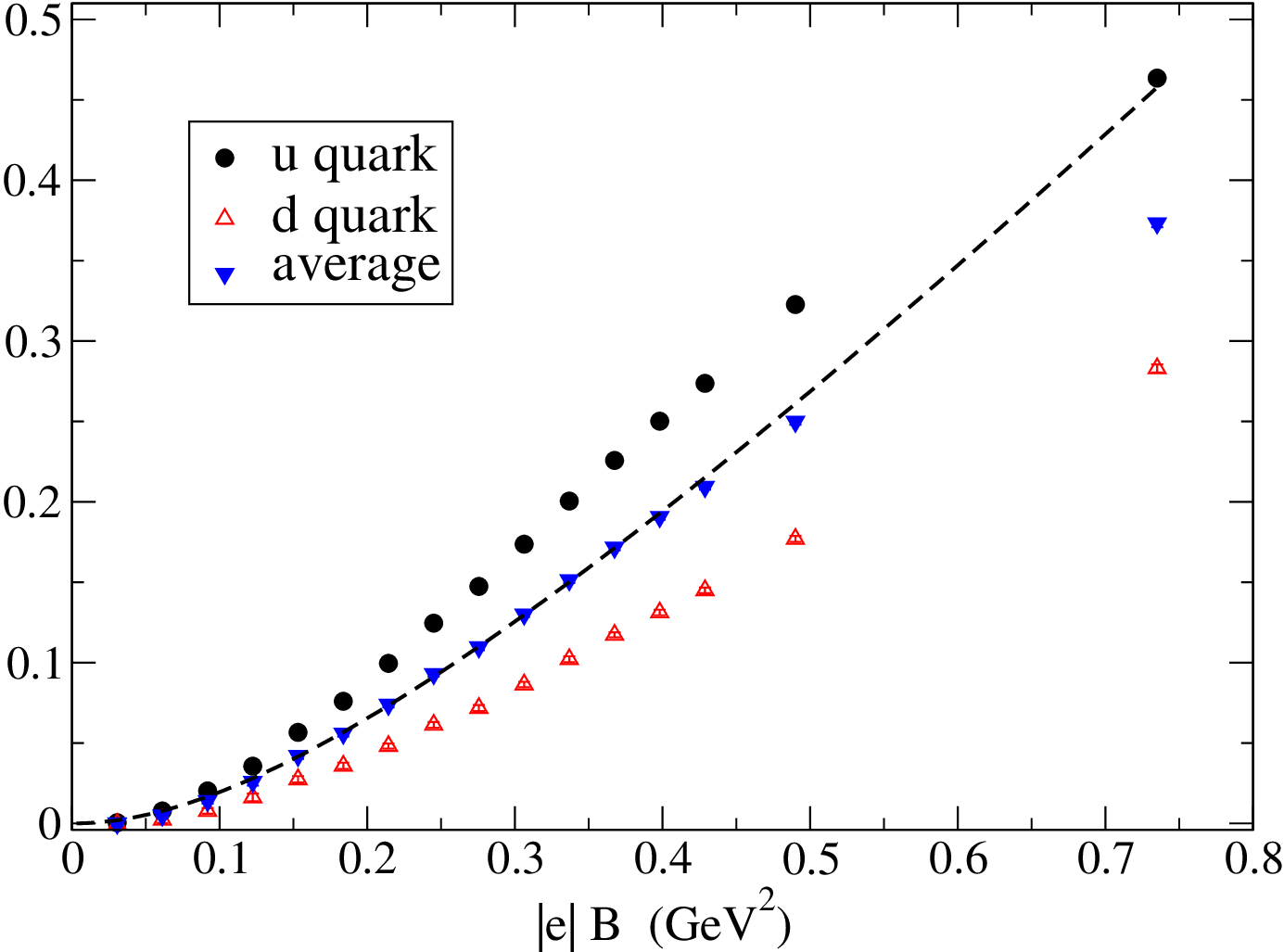}
\end{center}
\caption{
We show
the relative increment $r(B)$ (see Eq.~(\ref{rbdef})) as a function
of $|e| B$ for $N_f = 2$ QCD with $m_\pi \sim 200$ MeV.
 Data are reported separately for the 
$u$ and $d$ quark condensates, as well as for
their average, together with a best fit according to the functional
dependence inspired by chiral perturbation theory~\cite{cohen}.
Figure taken from~\cite{dene}. 
}
\label{fig:1}       
\end{figure}

In Ref.~\cite{dene}, where a standard staggered fermion and plaquette 
gauge discretization has been adopted, corresponding to a pseudo-Goldstone 
pion mass $m_\pi \sim 200$~MeV, an attempt has been made 
to separate the contributions to magnetic catalysis coming from 
the modification
of the Dirac operator from those coming from the modified gauge field distribution:
such contributions have been named "valence" and "dynamical", respectively.
One can show that such separation can be done 
consistently in the limit of small external fields. 
Indeed, let us define the quantities
\beq
\pbp^{val}_{u/d}(B) \equiv \int \mathcal{D}U {\cal P} [m,U,0]
{\rm Tr} \left( M^{-1}[m,B,q_{u/d}]\right)
\label{valence}
\eeq
\beq
\pbp^{dyn}_{u/d}(B) \equiv \int \mathcal{D}U {\cal P} [m,U,B]
{\rm Tr} \left( M^{-1}[m,0,q_{u/d}]\right) \, 
\label{dynamical}
\eeq
and, from those, the average (over flavor) quantities 
$\pbp^{val}$ and $\pbp^{dyn}$;
${\cal P}$ is the probability distribution for gauge fields
(including
quark loop effects) and $M$ is the fermion matrix.
In the first case, one looks at the spectrum of the fermion matrix which
includes the magnetic field explicitly, but is defined on 
non-Abelian configurations sampled with a measure 
$\mathcal{D}U {\cal P} [m,U,0]$ which, even 
including dynamical fermion contributions, is taken 
at $B = 0$ (partial quenching).
In the second case the measure term takes into account the external field, which 
is however neglected in the definition of the fermion matrix.
From $\pbp^{val}(B)$ and $\pbp^{dyn}(B)$ we can define the corresponding quantities $r^{val/dyn}$.

In the limit of small fields, $B$ acts as a perturbation
for both the measure term ${\cal P} [m,U,B]$ 
and the observable ${\rm Tr} \left( M^{-1}[m,B,q_{u/d}]\right)$:
assuming quadratic corrections in $B$ (however this 
is not an essential assumption), one
can write,
configuration by configuration:
\beq
{\cal P} [m,U,B] =  {\cal P} [m,U,0] + C\ B^2 + O(B^4)
\label{smallp}
\eeq
and
\beq
\hspace{-6pt} {\rm Tr} \left( M^{-1}[B]\right) = {\rm Tr} \left( M^{-1}[0]\right) + C'\ B^2 + O(B^4) \, 
\label{smallt}
\eeq
where $C$ and $C'$ depend in general on the quark mass and 
on the chosen configuration. 
Putting together the two expansions, one obtains~\cite{dene}
\beq
r(B) = r^{val}(B)  + r^{dyn}(B)  + O(B^4) \, ,
\label{additivity}
\eeq
therefore, at least in the limit of small fields, the separation of magnetic catalysis
in a valence and a dynamical contribution is a well defined concept.

\begin{figure}[t!]
$\, $
\vspace{0.9cm}
\begin{center}
\includegraphics[scale=.46]{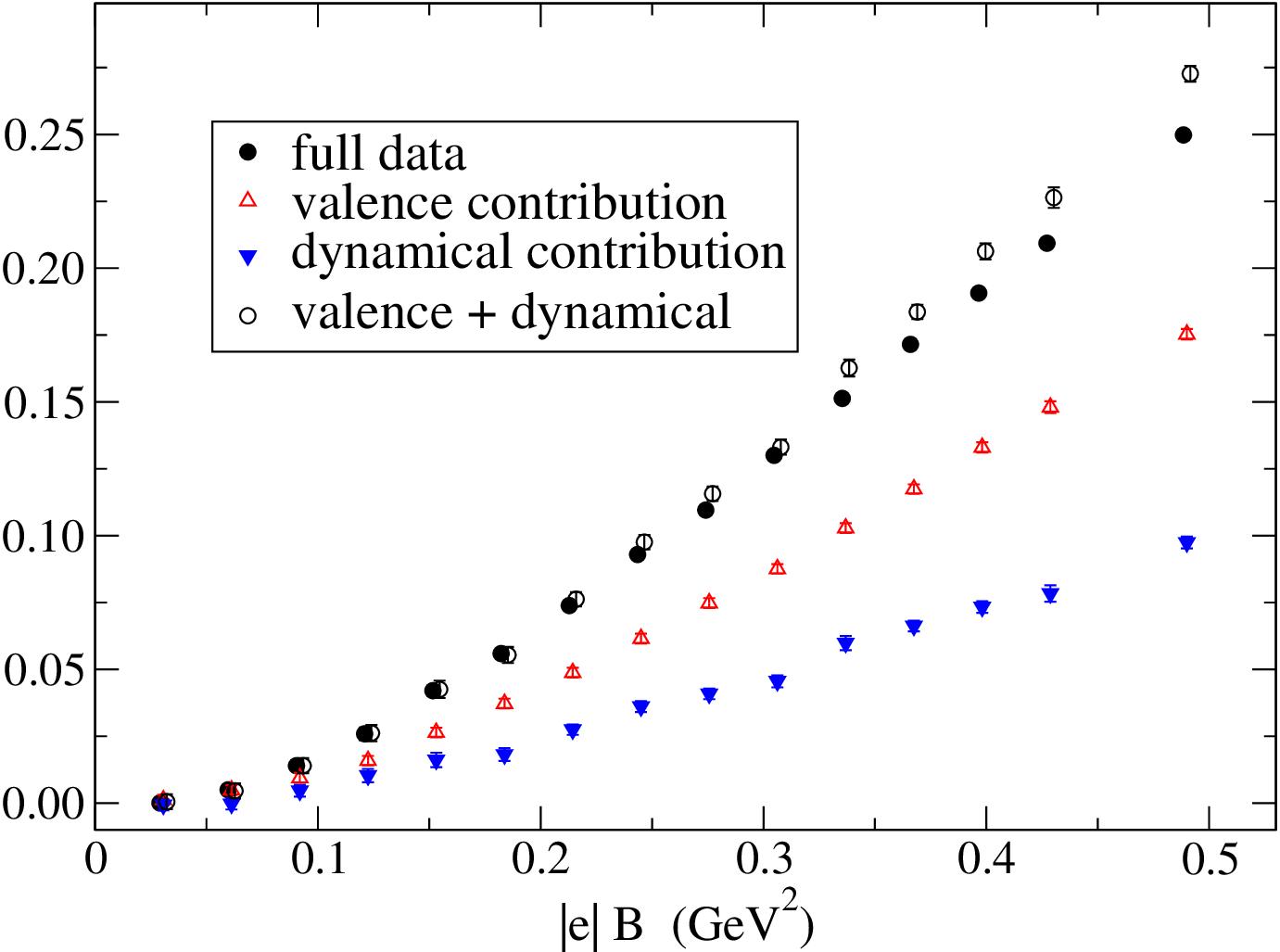}
\end{center}
\caption{
Relative increment of the average of the $u$ and $d$ quark condensates 
as a function of the magnetic field. $r(B), r^{val}(B), r^{dyn}(B)$
and $r^{val}(B) + r^{dyn}(B)$ are reported separately. From~\cite{dene}.
}
\label{fig:2}       
\end{figure}

In Fig.~\ref{fig:1} we report data obtained in~\cite{dene} for the 
relative increment of the $u$ and $d$ quark condensates and for their
average, as a function of $|e| B$. Magnetic catalysis is quite larger
for the $u$ quark with respect to the $d$ quark: this is expected on the basis
of the larger electric charge of the $u$ quark. Regarding their average, 
one observes that a functional dependence inspired by chiral perturbation
theory~\cite{cohen} fits well data, when taking into account the unphysical
quark mass spectrum considered in~\cite{dene}. A simple quadratic 
dependence describes well data in the regime of small background 
fields, approaching a linear behavior for larger fields,
while saturation effects
starts to be visible as $\sqrt{|e| B}$ reaches the scale of the UV cutoff,
which is quite low for the lattice setup used in~\cite{dene},
$a^{-1} \sim 0.7$ GeV.
Regarding the separation into valence and dynamical contributions, 
Fig.~\ref{fig:2}, taken again from~\cite{dene}, shows that it 
is indeed well defined in a significant range of magnetic fields,
where the two contributions are roughly additive, in the sense that 
their sum gives back the full signal, as expected at least 
in the limit of small fields. Moreover, from Fig.~\ref{fig:2}
one learns that the dynamical contribution is roughly 
40\% of the total signal, at least for the discretization 
and quark mass spectrum adopted in~\cite{dene}: 
this is an important contribution, which is larger than
other usual unquenching effects, which are  typically of the order of 20\%,
and reflects a significant modification in the distribution of gauge fields,
induced by the magnetic background field, which should be further 
investigated by future studies.

A new investigation of magnetic catalysis for the QCD vacuum
has been done recently~\cite{bali2}, making use of an improved 
gauge and (stout) rooted staggered discretization of the theory:
results for the 
average increment of the $u$ and $d$ quark condensates have been reported
for $N_f = 2 + 1$ flavors (i.e. including also strange quark 
contributions), adopting physical quark masses and after extrapolation to the continuum limit.
Also in this case one observes, for the relative increment $r(B)$ 
(see Fig.~1 of Ref.\cite{bali2}), a quadratic
dependence on $|e| B$ for small external fields, followed by an
almost linear dependence as $|e| B \gg m_\pi^2$; a nice
agreement is found with predictions from chiral perturbation theory
and from PNJL models~\cite{ruggieri}, at least for not too large fields.
It is interesting to notice that, in the regime of small fields,
the results of~\cite{dene} and those of~\cite{bali2} are compatible
with each other if they are rescaled by a factor
$m_\pi^2$ (which is different in the two studies)\footnote{The author
thanks G. Endrodi for giving him access to the continuum extrapolated 
data of~\cite{bali2}.
}:
the prediction from chiral perturbation theory is indeed,
in the limit of small fields~\cite{cohen,dene}:
\beq
\label{low_H}
r(B) \simeq \frac{(|e|B)^2}{96 \pi^2 F_\pi^2 m_\pi^2} 
\eeq

The authors of~\cite{ilgenfritz}, instead, have presented an 
investigation of magnetic catalysis for QCD with two colors and 
4 staggered flavors, which are degenerate both in mass and electric 
charge: that permits, contrary to Refs.~\cite{dene,bali2},
to avoid rooting and the possible systematic effects related to it.
The lattice discretization is standard, as in~\cite{dene}, with also
a similar range of pseudo-Goldstone pion masses. 
Results are qualitatively similar to those obtained in~\cite{dene,bali2}, however the authors try also 
an extrapolation to the chiral limit, thus verifying the 
prediction~\cite{smilga} for a linear dependence on $|e| B$ in such limit.

Till now we have discussed about magnetic catalysis in the QCD
vacuum, i.e. in the low temperature, deconfined phase, where a
consistent picture emerges, both from model and lattice computations.
The situation is less clear as one approaches the high temperature,
deconfined regime: results from~\cite{bali,bali2} have shown that
the growth of the chiral condensate with the magnetic field may
stop and even turn into an inverse magnetic catalysis at high enough 
temperatures; such effect may have a possible interpretation 
based on the effects of a strong magnetic field on gluodynamics
(see~\cite{anisotropic,shov,galilo} and the discussion in 
Section~\ref{gluodyn}). 
Inverse magnetic catalysis has been predicted by some model 
computation~\cite{schmitt}, however in a regime of low temperature
and high baryon density, which is different from that explored in~\cite{bali,bali2};
recently a possible explanation has been proposed according to which 
inverse catalysis derives from  dimensional reduction induced by the magnetic 
field on neutral pions~\cite{fuku}.
On the other
hand, the lattice studies reported in~\cite{demusa,ilgenfritz}
do not show such effect, reporting instead for standard magnetic 
catalysis at all explored temperatures. 
We will comment on   
the possible origin of such discrepancies
in the following Section, where we discuss about the effects
of strong external fields on the QCD phase diagram.

Finally, among the studies aimed at clarifying the effects of strong
magnetic fields on chiral dynamics, we mention those addressing 
the determination of the magnetic susceptibility of the chiral
condensate, both in quenched~\cite{itep11,itep2} and in
unquenched QCD~\cite{bali3}, which is part of the total contribution
to the magnetic susceptibility of the QCD vacuum. In particular,
recent unquenched results~\cite{bali3} show that such contribution
is of diamagnetic nature.

\section{QCD Phase diagram in External Fields}
\label{sec:phasediagram}

Similarly to what happens with other external parameters, like 
a finite baryon chemical potential, 
the introduction of a background field, either magnetic
of chromomagnetic, can modify the phase structure of QCD. 
The interest in such issue is theoretical, on one side, since 
a background field can be viewed as yet another parameter of the 
QCD phase diagram, which can help in getting a deeper understanding of the 
dynamics underlying deconfinement and chiral symmetry restoration.
There is however also great phenomenological interest, since 
strong background fields may be relevant to the cosmological
QCD transition, to heavy ion collisions and to the physics of some
compact astrophysical objects. The main questions regarding
the QCD transition that one
would like to answer can be summarized as follows:\\
1) do deconfinement and chiral symmetry restoration remain entangled,
or is a strong enough magnetic field capable of splitting the two 
transitions?\\
2) does the (pseudo)critical temperature depend on the background 
field strength, and how?\\
3) does the nature of the transition depend on the background field
strength?\\
4) does any new, unexpected phase of strongly interacting 
matter emerge for strong enough background fields?

Many model computations exist, which try to answer those 
questions~\cite{model1,model2,model3,model4,model5,model6,model7,model8,model9,model10,model11,model12,model13,model14,model15,model16,model17,model18,model19,model20}, in the following we will discuss results
based on lattice simulations. The focus will be on aspects regarding 
deconfinement and chiral symmetry restoration, leaving aside 
other issues, like the possible emergence 
of a superconductive phase for strong enough magnetic 
fields~\cite{superc,itep5,superc1,superc2}.

\begin{figure}[t!]
$\ $
\vspace{0.9cm}
\begin{center}
\includegraphics[scale=.46]{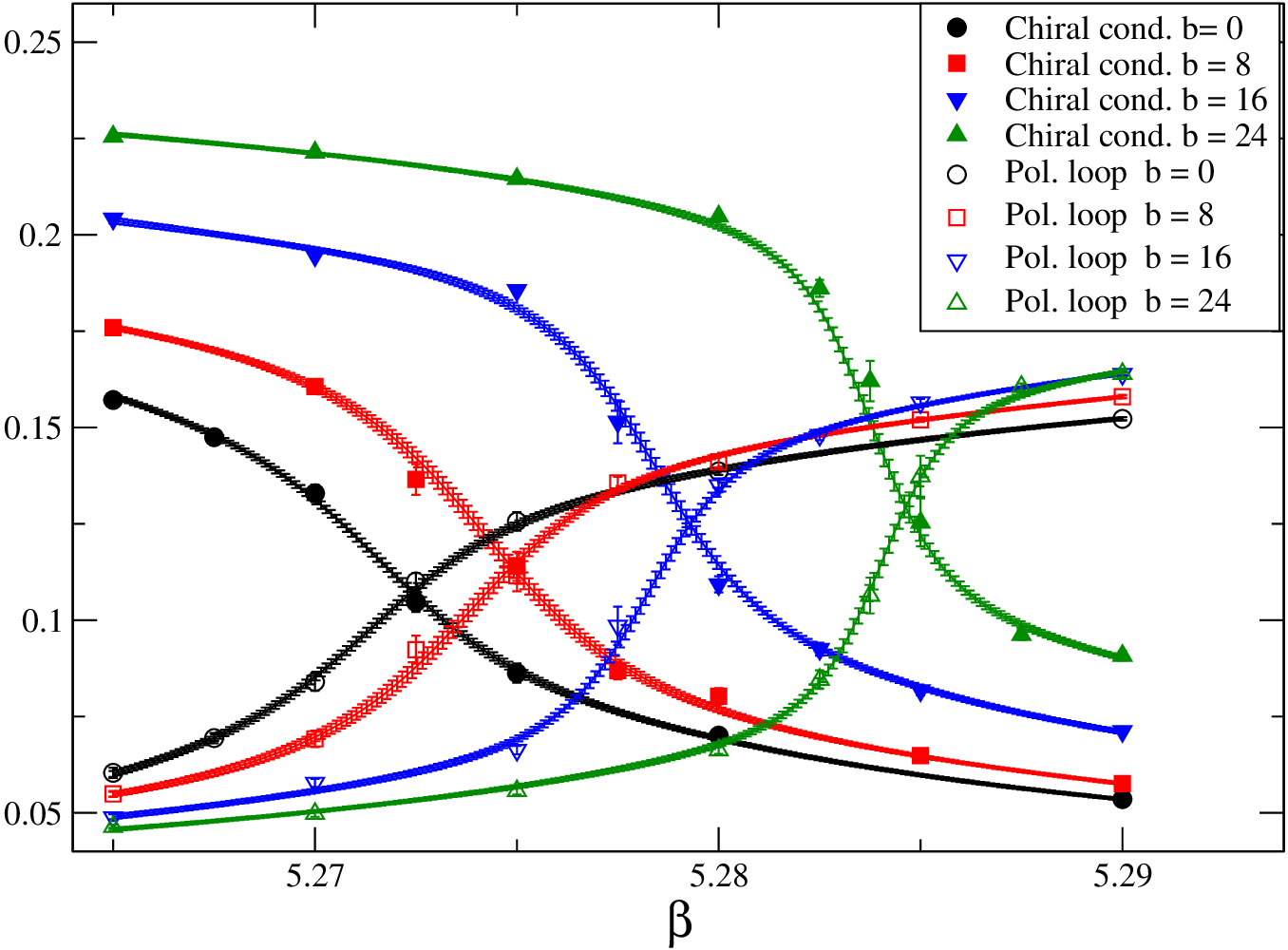}
\end{center}
\caption{Behavior of the chiral condensate and of the Polyakov loop,
as a function of the inverse gauge coupling $\beta$ and for 
different magnetic field quanta $b$, on a $16^3\times 4$ lattice and 
for a pion mass $m_\pi \sim 200$ MeV. Figure taken from~\cite{demusa}.
}
\label{fig:3}       
\end{figure}

\subsection{Deconfinement Transition in a Strong Magnetic Background}

Investigating the effects of a background magnetic field on 
thermodynamics and on the phase diagram of QCD necessarily requires an 
approach in which the presence of the magnetic field is taken into
account at the level of dynamical fermions.

A first study along these lines has been presented in~\cite{demusa},
where finite temperature QCD with two degenerate flavors has been simulated
in presence of a constant, uniform magnetic background.  A standard 
rooted staggered discretization (see Eq.~(\ref{partfun1}))
and a plaquette gauge action have been
adopted, with a lattice spacing of the order of 
0.3 fm, and different quark masses,
corresponding to a pseudo-Goldstone pion mass 
ranging from 200 MeV to 480 MeV, with 
magnetic fields 
going up to $|e|B\sim0.75$ GeV$^2$.

\begin{figure}[t!]
$\ $
\vspace{0.9cm}
\begin{center}
\includegraphics[scale=.46]{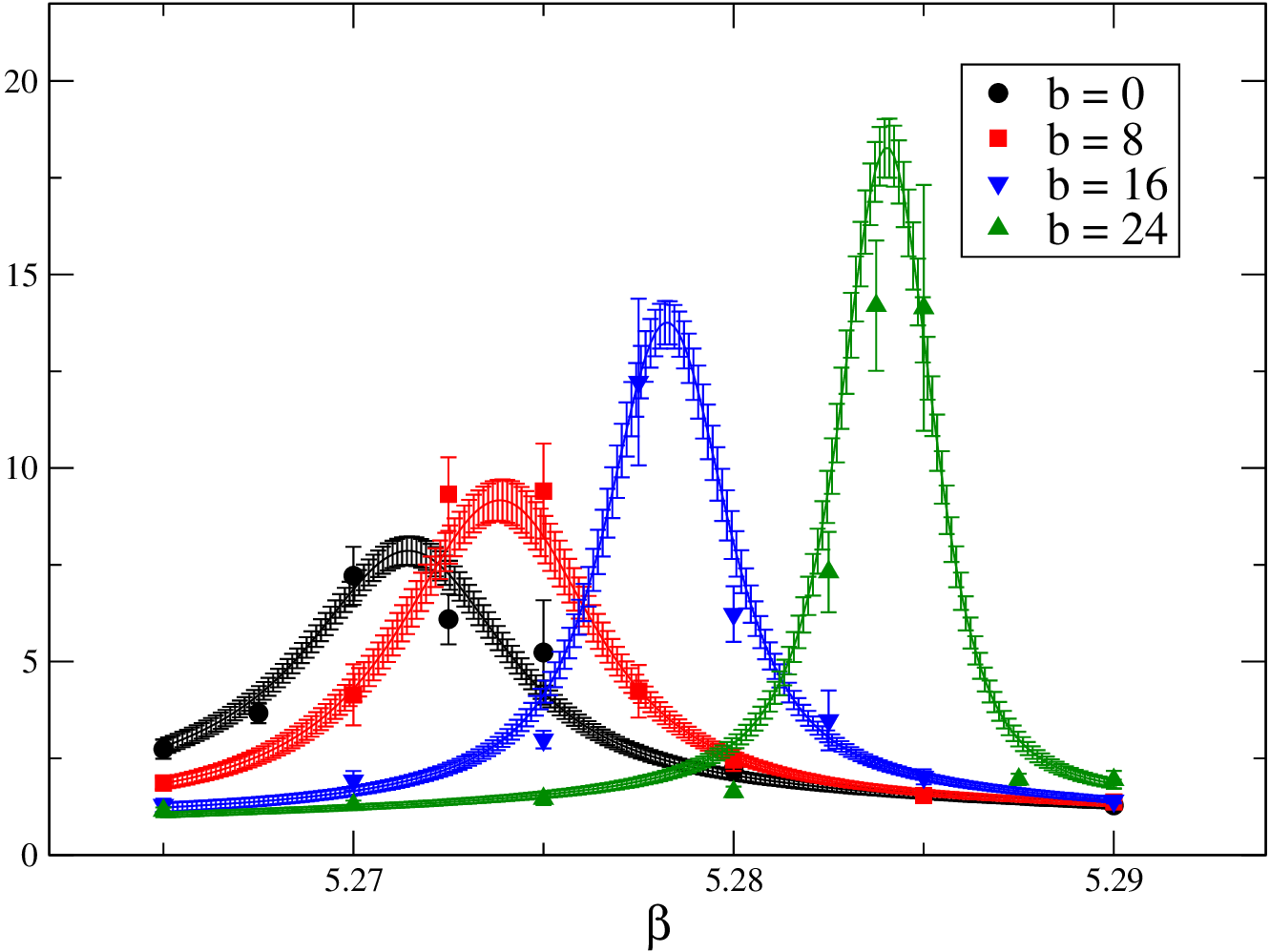}
\end{center}
\caption{
Disconnected susceptibility of the chiral condensate 
as a function of the inverse gauge coupling $\beta$ and for 
different magnetic field quanta $b$, on a $16^3\times 4$ lattice and 
for a pion mass $m_\pi \sim 200$ MeV. Figure taken from~\cite{demusa}.
}
\label{fig:4}       
\end{figure}

\begin{figure}[t!]
$\ $
\vspace{0.9cm}
\begin{center}
\includegraphics[scale=.46]{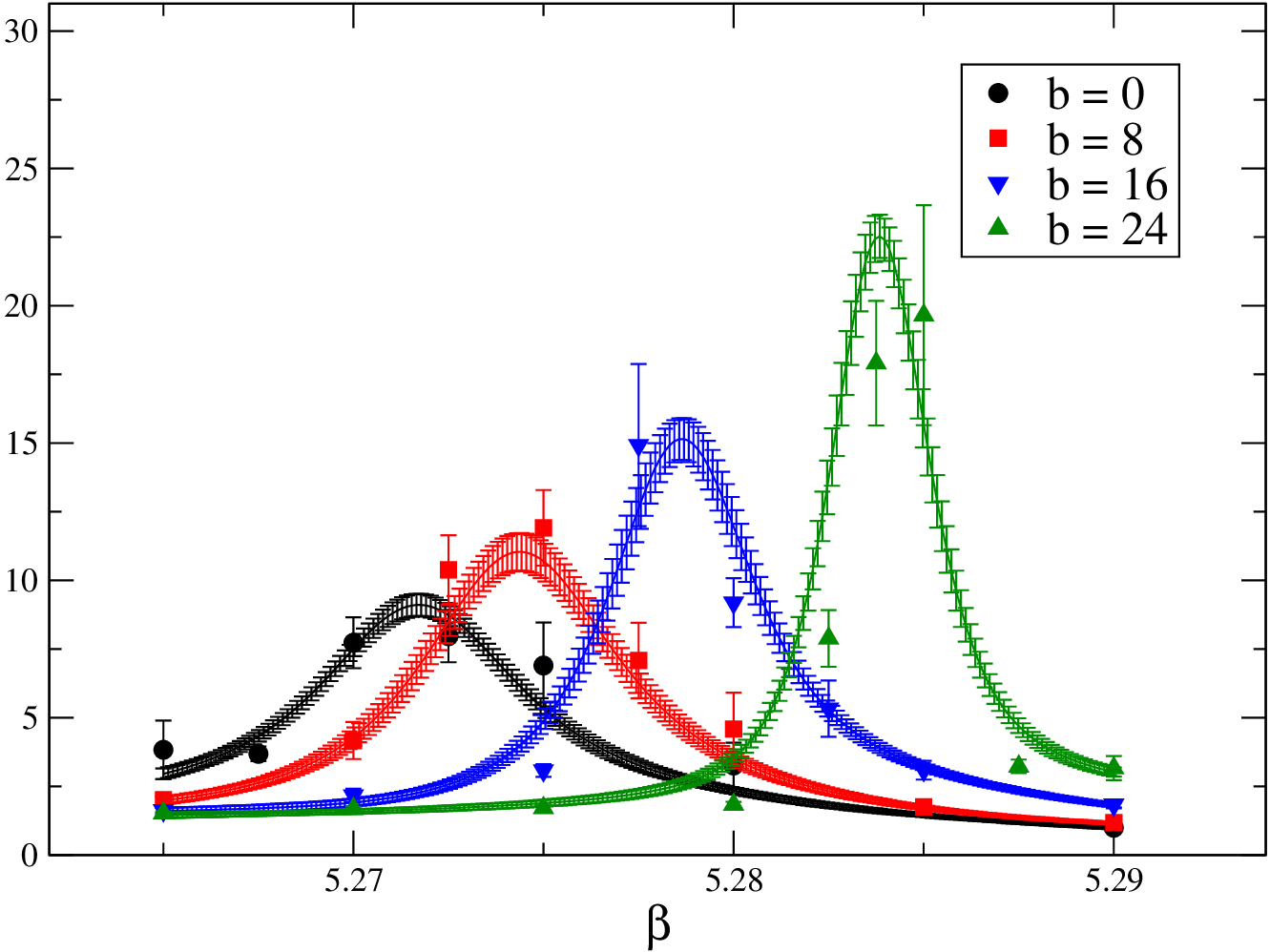}
\end{center}
\caption{
Polyakov loop susceptibility 
as a function of the inverse gauge coupling $\beta$ and for 
different magnetic field quanta $b$, on a $16^3\times 4$ lattice and 
for a pion mass $m_\pi \sim 200$ MeV. Figure taken from~\cite{demusa}.
}
\label{fig:4bis}       
\end{figure}

\begin{figure}[t!]
$\ $
\vspace{0.9cm}
\begin{center}
\includegraphics[scale=.46]{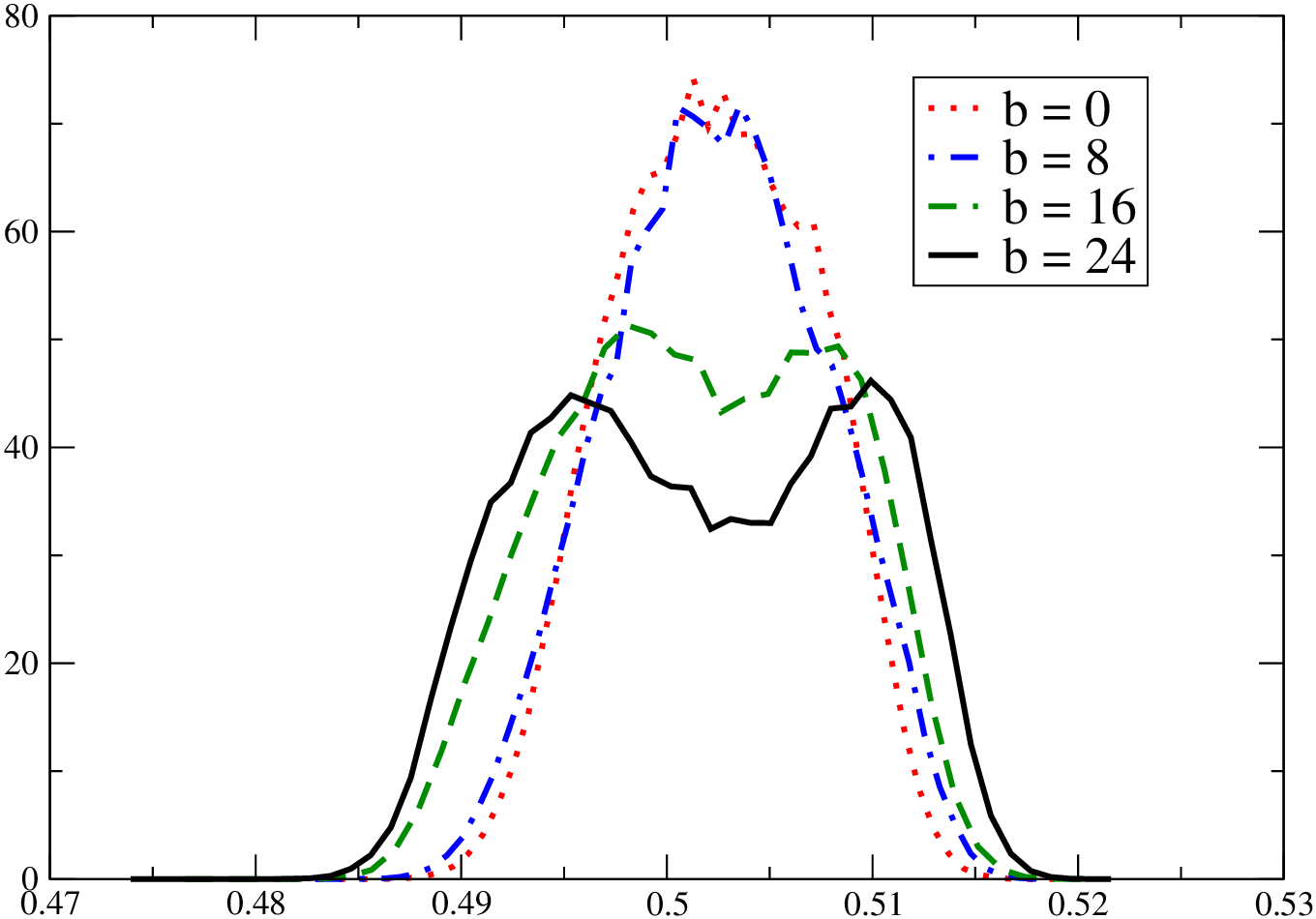}
\end{center}
\caption{
Plaquette distribution at the pseudocritical temperature
for $m_\pi \sim 200$ MeV
and for various different values of the magnetic field.
 Figure taken from~\cite{demusa}.
}
\label{fig:5}       
\end{figure}

Some results from~\cite{demusa} are reported in 
Figs.~\ref{fig:3}, \ref{fig:4}, \ref{fig:4bis} and \ref{fig:5}. In particular, 
Fig.~\ref{fig:3} shows the behavior of the chiral condensate and of the 
Polyakov loop as a function of the inverse gauge coupling $\beta$, 
for the lowest pion mass explored
($m_\pi \sim 200$ MeV) and for various magnetic fields (expressed
in units $b$ of the minimum quantum allowed by the periodic boundary 
conditions).
Remember that
the physical temperature, $T = 1/L_t a(\beta)$, is a monotonic, increasing 
function of $\beta$, and that, on the $16^3 \times 4$ lattice explored
in~\cite{demusa}, the magnetic field is given by
$|e| B = (3 \pi T^2/8)\, b \sim (0.03\ {\rm GeV}^2)\, b\ $ around the transition.

Three facts are evident from Fig.~\ref{fig:3}. The chiral condensate
increases, as a function of $B$, for all explored temperatures.
The inflection points 
of the chiral condensate and of the Polyakov loop, signalling
the location of the pseudo-critical temperature, 
move together towards
higher temperatures as the magnetic field increases, meaning 
that deconfinement
and chiral symmetry restoration do not disentangle, at least in the
explored range of external fields. The drop (rise) of the chiral condensate
(Polyakov loop) at the transition seems sharper and sharper 
as the magnetic field increases, meaning that the (pseudo)transition 
is strengthening.

Such facts are confirmed by Figs.~\ref{fig:4} and \ref{fig:4bis}, where
the susceptibilities of the chiral condensate and of the Polyakov 
loop are reported, for different values of the magnetic field: 
the peaks move to higher $\beta$ (hence to higher $T$) and become sharper 
and sharper as $B$ increases, i.e. their height increases and their 
width decreases. The increased strength of the transition
is also appreciable from the plaquette (pure gauge action) distribution 
at the pseudocritical coupling, which is reported in Fig.~\ref{fig:5}
and which seems to evolve towards a double 
peak distribution, typical of a first order transition, as 
$B$ increases. Such hints for a change in the nature
of the transition, however, have not yet been confirmed by simulations
on larger lattices. Regarding the increase in pseudo-critical temperature,
we notice that it is quite modest in magnitude and of the order
of $2$\% at $|e| B \sim$ 1 GeV$^2$.
\\

After the exploratory study of Ref.~\cite{demusa}, new studies 
have appeared in the literature, which have added essential 
information and also opened new interesting questions. Let us first consider
the investigation reported in~\cite{bali}. There are three 
essential differences, with respect to~\cite{demusa}, in the 
lattice discretization and approximation of QCD:\\
1) still within a rooted staggered fermion formulation, 
improved gauge (tree level Symanzik improved) and fermionic 
(stout link) discretizations have been implemented, and
different lattice spacings have been explored, 
in order to get control over the continuum limit;\\
2) the authors have explored $N_f = 2+1$ QCD, i.e. they have considered 
strange quark effects;\\
3) a physical quark mass spectrum has been adopted.

Concerning results, instead, the most striking difference with respect 
to~\cite{demusa} is that the pseudocritical temperature decreases,
instead of increasing, in presence of the magnetic background,
in particular it becomes 10-20\% lower for
 $|e| B \sim$ 1 GeV$^2$. 
The authors of~\cite{bali} have put
the decrease of the critical temperature in connection
with another unexpected phenomenon
that they observe, 
namely  inverse magnetic catalysis, i.e. the fact that,
at high enough temperatures,  
the chiral condensate starts decreasing, instead
of increasing, as a function of $B$. A possibility pointed out in~\cite{bali2}
is that inverse catalysis may originate from the gluonic sector, i.e. 
the distribution of gluon fields may change, as an indirect effect
of the magnetic field mediated by quark loops, so as to destroy
the chiral condensate. We point out that the study of~\cite{dene}
has shown instead that, at least at zero temperature, the modified distribution
of gluon fields contributes to increase magnetic catalysis 
(dynamical contribution), however the situation may indeed be quite different
around and above the deconfinement temperature.

Other aspects pointed out in~\cite{demusa}
have instead been confirmed by~\cite{bali}: 
in particular the absence of a clear splitting between 
deconfinement and chiral symmetry restoration, induced by
the external field, and an 
increased strength of the transition at $B \neq 0$, even without evidence for a change
of its nature.
\\

A different approach has been followed by the authors of 
Ref.~\cite{ilgenfritz}: they have explored 2 color QCD, making use
of a discretization very close to that adopted in~\cite{demusa}, i.e.
with the same gauge and staggered action and a similar range of pion masses.
However, to avoid the possible systematic effects connected with 
taking the square or fourth root of the fermionic determinant, they 
have considered a theory with four degenerate
flavors, $N_f = 4$, all carrying the same mass and electric charge.
Their results are in quite good agreement with those of~\cite{demusa}: deconfinement and chiral symmetry restoration
do not split; the transition gets sharper; the pseudo-critical
temperature increases as a function of $B$ and no inverse catalysis
is observed, i.e. the chiral condensate is an increasing function of 
$B$ for all explored temperatures. The increase in $T_c$ is also
larger than what observed in~\cite{demusa},
being of the order
of $10$\% at $|e| B \sim$ GeV$^2$: that can possibly be explained
by the fact that the authors of Ref.~\cite{ilgenfritz} make use
of $4$ flavors, instead of 2, moreover all carrying the 
charge of the $u$ quark, hence the influence 
of the magnetic field on the system can be larger.

Another interesting aspect, explored in~\cite{ilgenfritz}, regards
the actual fate of chiral symmetry above the transition and in presence of 
the background field. The authors have explicitly verified, by performing
an extrapolation to the chiral limit, that the chiral condensate vanishes
in the high temperature phase, 
i.e. that chiral symmetry is 
exact above $T_c$,
also in presence of the magnetic backgroud. 
It is interesting to notice that, in the system 
explored in~\cite{ilgenfritz}, the magnetic
field does not break any of the flavor symmetries of the 
theory, since all quarks carry the same electric charge. However
also in the standard case, in which quarks carry different charges, there
are still diagonal chiral flavor generators which commute with the 
electric charge matrix and are not broken
by the introduction of an external electromagnetic background field,
so that an investigation similar to that performed in~\cite{ilgenfritz}
should be done, to enquire about 
the actual realization of the unbroken symmetries above $T_c$.
\\

To summarize, present lattice investigations about 
the influence of a strong
magnetic field on the QCD phase diagram have given 
consistent indications about two facts: a magnetic field 
increases the strength of the QCD transition, however it does not change
its nature, at least for $|e| B$ up to 1 GeV$^2$; in
the same range of external fields, no significant splitting of 
deconfinement and chiral symmetry restoration has been detected.

There is a controversial issue, instead, regarding the location of 
the pseudocritical temperature, $T_c$, which increases as a function 
of $|e| B$ according to the results of~\cite{demusa,ilgenfritz},
while it decreases according to the results of~\cite{bali}.
It is perfectly possible that various
improvements regarding the lattice discretization may turn the 
very slowly increasing function $T_c(B)$ determined in~\cite{demusa}
into a decreasing function. The main point is to understand 
which aspect is more directly related to the change of behavior
and to the appearance of inverse catalysis. Is it just a problem 
of lattice discretization and approach to the continuum limit, or
can the difference be traced back to the larger pion mass used in~\cite{demusa}? 
Could instead the introduction of the strange quark
explain the differences? 
We believe that a clear answer to those questions can be given
by considering each single aspect separately, and that the answer 
will be important by itself, since it will help clarifying
the origin of unexpected behaviors such as inverse magnetic catalysis.

\subsection{Deconfinement Transition in a Chromomagnetic Background}

\begin{figure}[t!]
$\ $
\vspace{1.0cm}
\begin{center}
\includegraphics[scale=0.46]{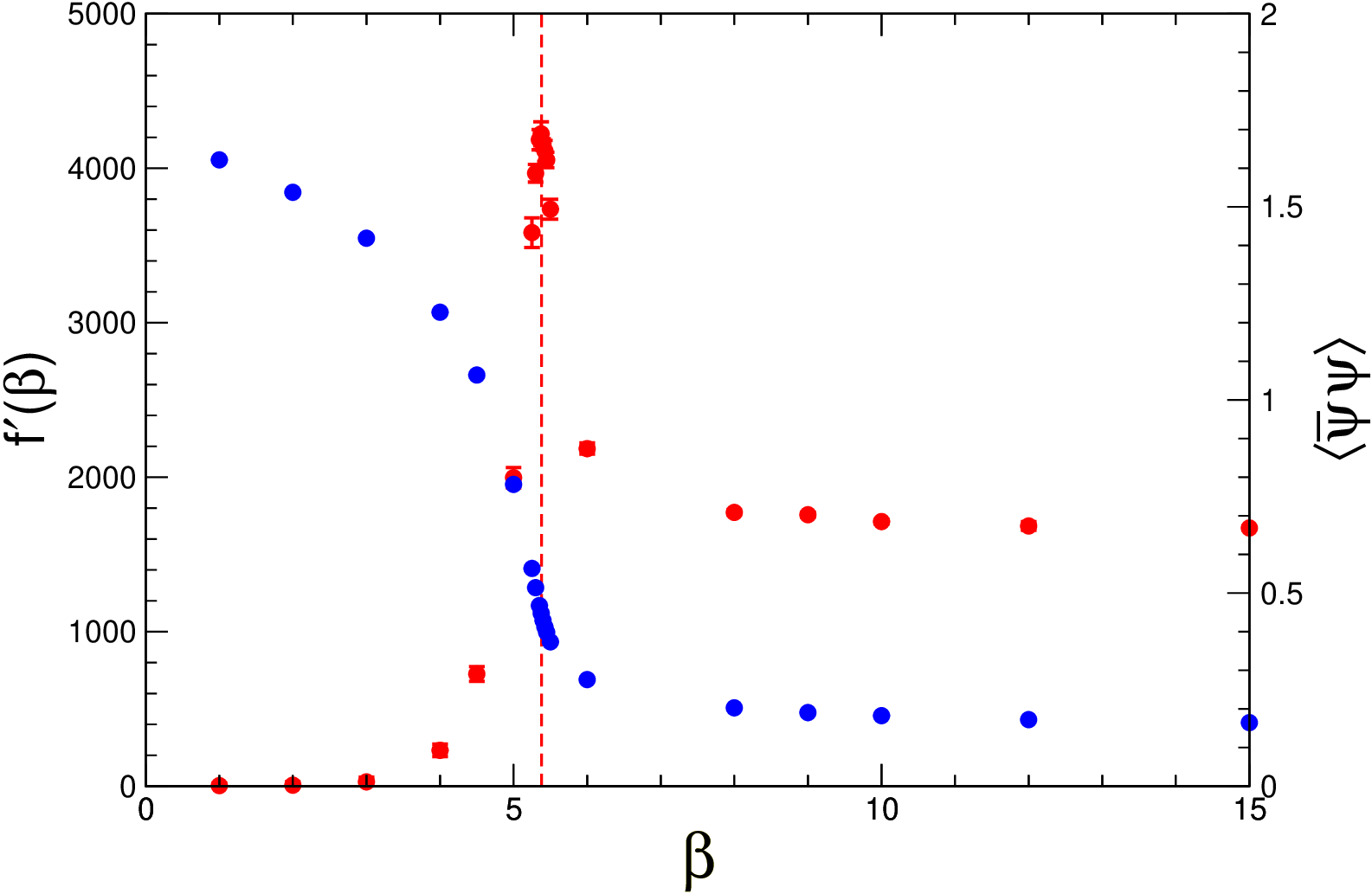}
\end{center}
\caption{
Peak of the derivative of the free energy of the background field
with respect to the inverse gauge coupling, 
together with the chiral 
condensate,  for $N_f = 2$ QCD
and for a background chromomagnetic field 
$g B = \pi/(16 a^2) \sim 0.35$ GeV$^2$. Figure taken from~\cite{cea_full}.
}
\label{fig:6}       
\end{figure}

\begin{figure}[t!]
$\ $
\vspace{1.0cm}
\begin{center}
\includegraphics[scale=0.46]{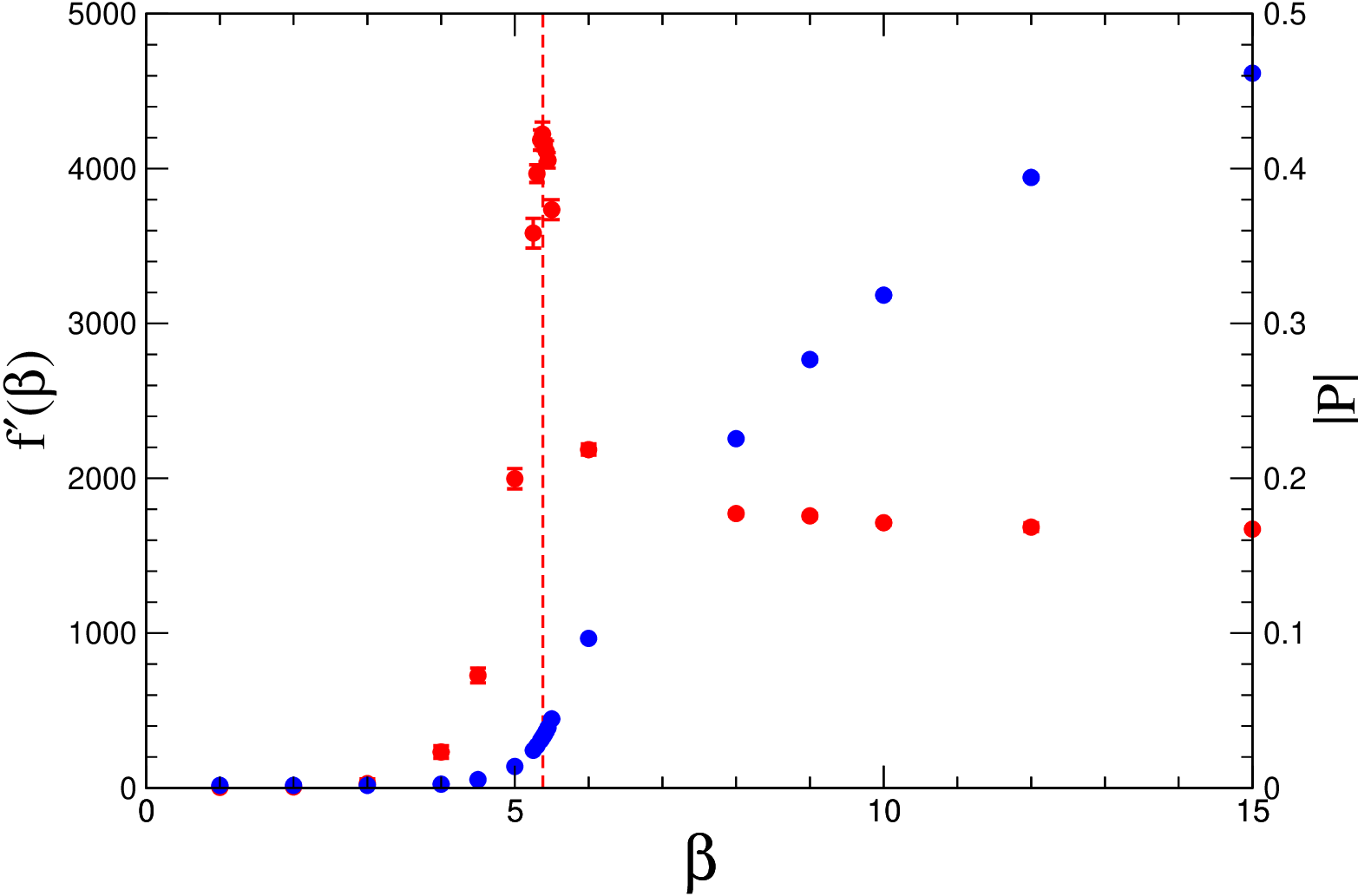}
\end{center}
\caption{
Peak of the derivative of the free energy of the background field
with respect to the inverse gauge coupling, 
the Polyakov loop,  for $N_f = 2$ QCD
and for a background chromomagnetic field 
$g B = \pi/(16 a^2) \sim 0.35$ GeV$^2$. Figure taken from~\cite{cea_full}.
}
\label{fig:6bis}       
\end{figure}

A chromomagnetic background field influences directly gluodynamics,
hence it makes sense to investigate its effects also in pure 
gauge simulations. A standard approach to introduce colored
background fields has been described in Section~\ref{subsec:3} 
and is based on the formalism of the lattice 
Schr\"odinger functional~\cite{sch_1,sch_2}: such approach has been 
adopted in the literature to study the influence of background 
fields both in pure Yang-Mills theories and in full QCD,
and for various kinds of background fields, going from those
corresponding to a uniform magnetic field to those produced 
by magnetic monopoles~\cite{cea_old,cea_polosa,cea_quench_1,cea_quench_2, cea_full, antropov,
cea_monofull}. In the following we shall consider the case of a constant
background field~\cite{cea_quench_1,cea_quench_2,cea_full}.

Like a magnetic field, also a chromomagnetic background field 
leads to a shift of both chiral symmetry restoration,
signalled by the drop of the chiral condensate, and of deconfinement,
signalled by the rise of the Polyakov loop.
The two transitions move together, moreover it is interesting 
to notice that the transition point coincides with the temperature
at which the free energy (effective potential) of the background
field shows a sudden change, the background field being screened
(not screened) in the confined (deconfined) phase.

As an example, in Figs.~\ref{fig:6} and \ref{fig:6bis}, taken from~\cite{cea_full},
we show the behavior of the chiral condensate and of the Polyakov loop
versus the inverse gauge coupling, $\beta$, together with the
derivative of the free energy with respect to $\beta$, $f'(\beta)$,
which shows a sharp peak at the same point at which chiral
symmetry restoration and deconfinement take place.
Results make reference to simulations of $N_f = 2$ QCD
on a $32^3 \times 8$ lattice, with a 
standard pure gauge and staggered fermion formulation, 
a lattice spacing $a \simeq 0.15$ fm 
and a bare quark mass $am  = 0.075$; the value
of the magnetic background field is $g B = \pi/(16 a^2) \sim 0.35$ GeV$^2$.

\begin{figure}[t!]
$\ $
\vspace{0.9cm}
\begin{center}
\includegraphics[scale=.46]{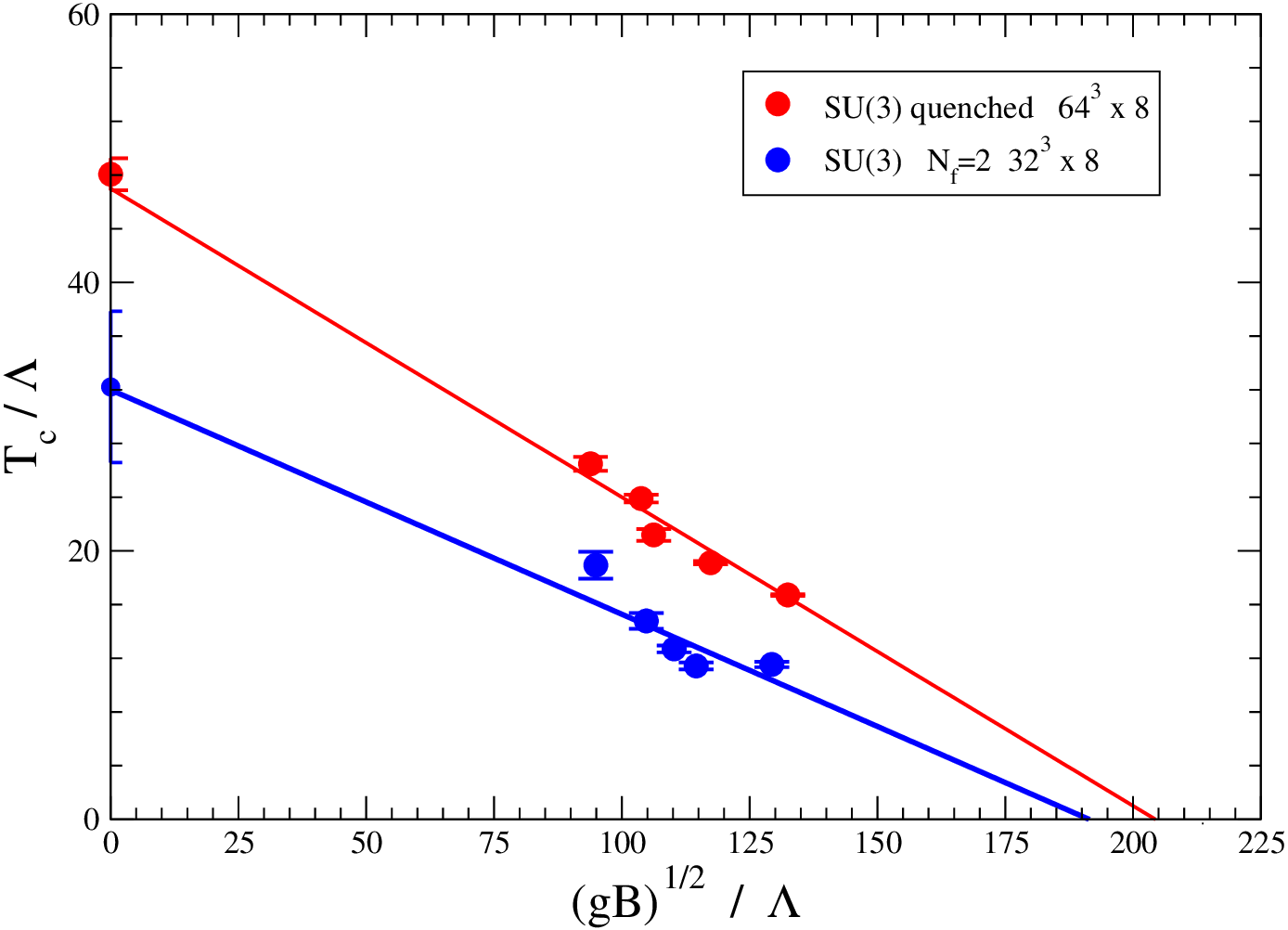}
\end{center}
\caption{
Dependence of the critical temperature $T_c$ on the external 
chromomagnetic field for the pure gauge theory and for 
$N_f = 2$ QCD. Physical quantities are expressed in terms of 
the parameter $\Lambda \sim 6$ MeV. 
}
\label{fig:7}       
\end{figure}

Present lattice results indicate that, both for pure gauge and full QCD,
the transition temperature decreases in presence of a constant
background chromomagnetic field. This is shown in Fig.~\ref{fig:7}~\cite{cea_full},
where the critical temperature $T_c$ (expressed in units of the 
parameter $\Lambda \sim 6 $ MeV), 
is shown as a function of $gB$, both for the pure
gauge theory and $N_f = 2$ QCD. The change in the critical temperature
is, in general, much larger than what happens in presence of 
an electromagnetic background: this can be interpreted in terms
of the fact that a colored background field directly affects gluodynamics.
An extrapolation of lattice results would even hint, as shown
in Fig.~\ref{fig:7}, at the presence of a zero temperature deconfining
transition, for $\sqrt{g B}$ of the order of 1 GeV.
It would be interesting to investigate this possibility further,
in the future,
for the possible cosmological and astrophysical implications it could 
have. 
One should repeat the study of~\cite{cea_full} in presence 
of more physical quark masses (with the values adopted in~\cite{cea_full}, $m_\pi$ is of the order of 400-500 MeV)
and closer to the continuum limit, also 
in order to reach higher values of the external field.

\begin{figure}[t!]
$\ $
\vspace{0.9cm}
\begin{center}
\includegraphics[scale=.46]{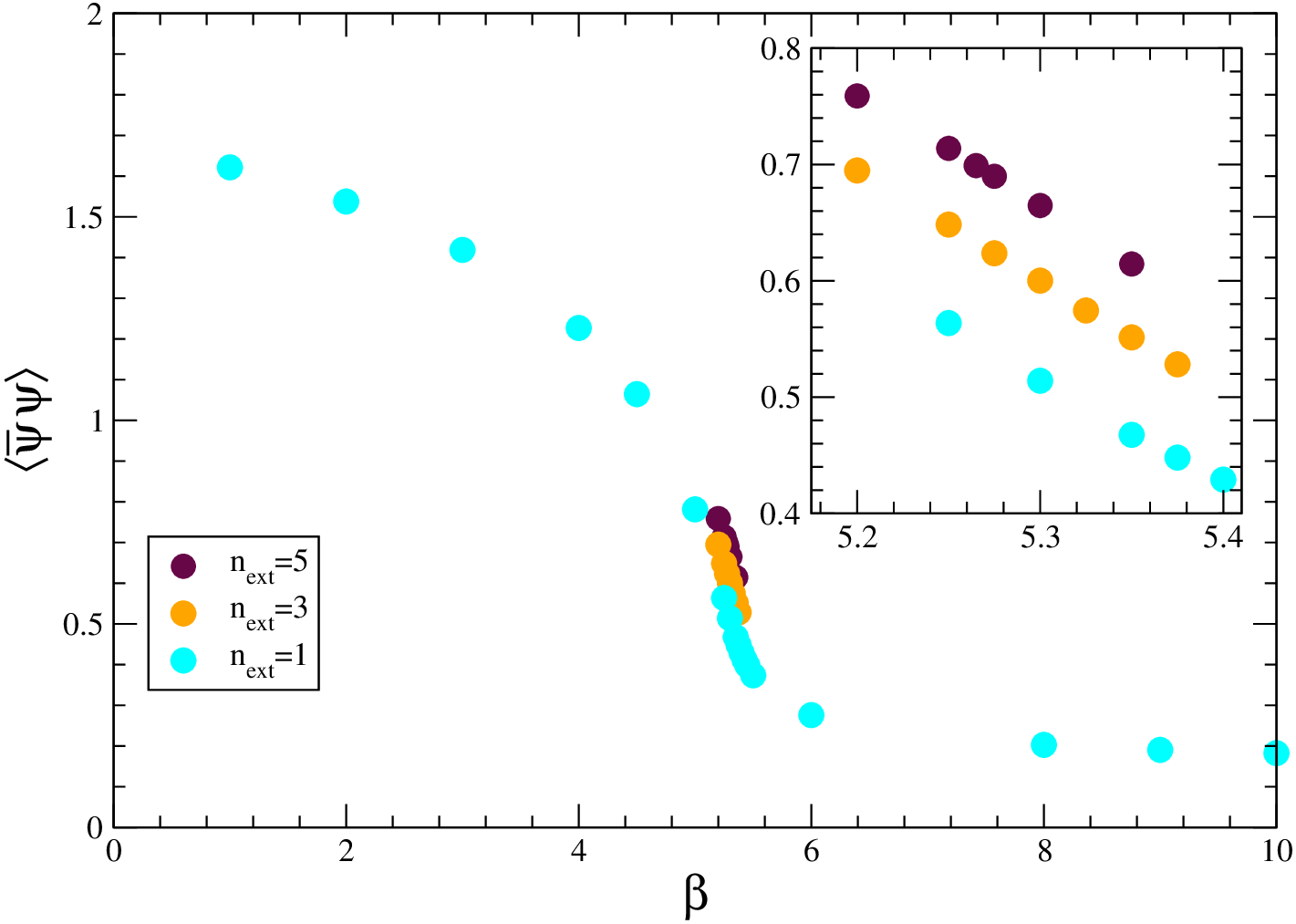}
\end{center}
\caption{
Chiral condensate as a function of the inverse gauge coupling $\beta$ for 
different values of the external chromomagnetic field 
$gB = \pi n_{\rm ext} /(16 a^2)$ and for $N_f = 2$ QCD. Figure taken
from~\cite{cea_full}.
}
\label{fig:8}       
\end{figure}

As a final comment, we notice that~\cite{cea_full} also reports evidence about
magnetic catalysis induced by the chromomagnetic field. 
Fig.~\ref{fig:8}, in particular, shows the behavior of the chiral condensate
for a few values of $gB$: it is clear that, at least around the transition,
where data for all external fields are available, the chiral 
condensate grows as a function of $gB$. It is interesting to
notice that in this case $T_c$ decreases anyway,
even in presence of normal catalysis, i.e. inverse catalysis does not 
seem to be 
a necessary condition for a decreasing critical temperature, at least
in presence of finite quark masses.

\section{More on Gauge Field Modifications in External Electromagnetic Fields}
\label{gluodyn}

The lattice results that we have discussed in the previous Sections
show that an electromagnetic background field can have a strong 
influence not only on quark dynamics, but also on gluodynamics, 
even if the interaction is not direct
but mediated by quark loop effects. 
That does not come completely unexpected, at least in the strongly
interacting, non-perturbative regime.
Let us summarize a few facts:\\
1) we have shown that large part of magnetic catalysis is due to 
the modification of the gauge field distribution (dynamical contribution), 
i.e. that the increment of chiral symmetry breaking is lower by about 
40\% if gauge field configurations are sampled without taking
into account the background field~\cite{dene};\\
2) it is known, from model computations, that a strong 
electromagnetic background can modify gluon screening properties
and influence confinement dynamics, reducing the 
confinement scale~\cite{anisotropic,galilo,shov}: such effects
on gluodynamics have been proposed in~\cite{bali2} as a possible
explanation for inverse catalyis;\\
3) lattice computations have also explicitly shown that quantities
related to quark confinement, like the Polyakov loop, are influenced
by a background electromagnetic field, not only around the transition
but also well inside the confined and the deconfined phases (see
as an example the data reported in Fig.~\ref{fig:2});\\
4) the results of~\cite{ilgenfritz} have shown that the magnetic 
field can also induce an asymmetry in the non-Abelian plaquette 
values, especially around and above the transition,
with a possible significant effect on the equation of state of 
the Quark-Gluon Plasma.

It is surely of great importance to study and understand 
these effects in a deeper and more systematic way.
Is it possible, for instance, that electromagnetic background
fields may induce chromoelectric or chromomagnetic background fields?
That has been discussed in some recent model studies 
(see e.g.~\cite{galilo}) and
could have important phenomenological consequences, both for 
cosmology and astrophysics and for heavy ion collisions. 
In the following
we would like to discuss how 
lattice simulations could further contribute to the issue, with 
an accent on symmetry aspects.

Let us consider a uniform and constant magnetic background field. 
In order to understand some of the gluonic field modifications induced by it,
it is interesting to observe that it breaks explicitly
charge conjugation symmetry, and to ask how such breaking 
propagates to the gluon sector.

Charge conjugation on gluon fields means $A_\mu \to - A_\mu$ or,
at the level of lattice gauge link variables, 
$U_\mu \to U_\mu^* $. Symmetry under it means that a gauge configuration
$U_\mu (x)$ has the same path integral probability of its 
complex conjugate $U_\mu^* (x)$:
\beq
\mathcal{D}U {\cal P} [U] = \mathcal{D}U^* {\cal P} [U^*] \, .
\eeq
As a consequence, all gluonic gauge inviarant quantities, i.e. traces
over closed parallel transports $W_C [U]$, including plaquettes
and Wilson loops, must be real:
\beq
\langle W_C [U] \rangle^* &=&  
\langle W_C^* [U] \rangle =
  \langle W_C [U^*] \rangle =
\int \mathcal{D}U {\cal P} [U] W_C [U^*] \nonumber \\ &=& 
\int \mathcal{D}U^* {\cal P} [U^*] W_C [U] = 
\int \mathcal{D}U {\cal P} [U] W_C [U] = 
\langle W_C [U] \rangle
\label{real_loop}
\eeq
where the first equality holds if the path integral measure is real, 
the second express the fact that 
$W_C$ is a trace over a product of links,
while the fourth is simply a change of integration variables. 
Eq.~(\ref{real_loop}) may be violated by 
possible spontaneous symmetry breaking effects, which may affect
some particular loops, as it happens for the
 Polyakov loop in the deconfined phase.

In presence of an electromagnetic background field, such 
property must be lost, since the breaking of charge conjugation symmetry
will propagate from the quark to the gluonic sector:
${\cal P} [U] \neq {\cal P} [U^*]$ because of the contribution
from the fermion determinant. Let us see that more explicitly,
considering the particular case of the trace over a plaquette.

The most direct way to look for effective quark contributions to 
the gauge field distribution is to consider the loop expansion of the 
fermionic determinant. If we write the fermion matrix as
$M = m_f {\rm Id} + D$, where $D$ is the discretization of the 
Dirac operator and $m_f$ is the bare quark mass, then the following 
formal expansion holds:
\beq
\det M = \exp \left( {\rm Tr} \log (m_f {\rm Id} + D) \right)
\propto \exp \left( - \sum_k \frac{(-1)^k}{k\, m_f^k} 
{\rm Tr}\, D^k  \right) \; .
\label{loopexp}
\eeq
$D^k$ is made up of parallel transports connecting lattice sites; 
the trace operator implies that only closed parallel transports 
will contribute. Moreover, for each contributing loop, there is 
an equal contribution from its hermitian conjugate.  
If $D$ is a standard, nearest neighbour discretization, the first non-trivial
term comes for $k = 4$
and contains a coupling to the 
plaquette operator, which can be expressed  as:
\beq
\frac{\Delta \beta (m_f)}{N_c} \frac{1}{2} \left( {\rm Tr}\, 
\Pi_{\mu\nu} (x)  + 
{\rm Tr}\, \Pi^\dagger_{\mu\nu} (x) \right) =
\frac{\Delta \beta (m_f)}{N_c}\, {\rm Re} {\rm Tr}\, \Pi_{\mu\nu} (x)
\label{realtrace}
\eeq
where $\Delta \beta (m_f)$ is a coefficient which
depends on $m_f$ and on the particular
fermion discretization adopted. The total effect can be viewed as
a simple renormalization of the inverse bare gauge coupling, $\beta \to 
\beta + \sum_f \Delta \beta (m_f)$, where the sum runs over flavors.

Let us now consider the effect of an external electromagnetic field. That
will change the definition of $D$, see e.g. Eq.~(\ref{covder}), 
so that each loop contribution to the determinant will get
a $U(1)$ phase from the external field. In particular, the expression
in Eq.~(\ref{realtrace}) for plaquettes
will be modified into
\beq
&&
\frac{\Delta \beta (m_f)}{N_c} \frac{1}{2} 
\left( e^{\I\, \phi_{\mu\nu}(x)} {\rm Tr} 
\Pi_{\mu\nu} (x)  + e^{- \I\, \phi_{\mu\nu}(x)} 
{\rm Tr} \Pi^\dagger_{\mu\nu} (x) \right) = \nonumber \\ &=&
\frac{\Delta \beta (m_f)}{N_c} \left( 
\cos(\phi_{\mu\nu}(x)) 
{\rm Re} {\rm Tr}\, \Pi_{\mu\nu} (x) - 
\sin(\phi_{\mu\nu}(x)) 
{\rm Im} {\rm Tr}\, \Pi_{\mu\nu} (x) \right)\, .
\label{imtrace}
\eeq
The phases $\phi_{\mu\nu}$ are in general non-trivial. For the particular
case of a uniform magnetic field $B$ in the $\hat z$ direction we have
$\phi_{xy} = q_f B a^2$, where $q_f$ is the quark charge.
This implies that, apart from the case in which there is exact 
cancellation among quark flavors carrying equal masses and opposite
electric charges (but this does not happen in real QCD), there
will be a non-trivial coupling to the imaginary part of some plaquette
traces,
which will then develop a non-zero expectation value. Notice that 
this is exactly what happens also for the Polyakov loop in presence
of an imaginary baryon chemical potential.

It is interesting to understand the meaning of a non-zero expectation
value for the imaginary part of the trace of the 
plaquette, in terms of continuum
quantities. Remember that, in the formal continuum limit,
the plaquette operator is linked to the gauge field
strenght $G_{\mu\nu} = G^a_{\mu\nu} T^a$ as follows:
\beq
\Pi_{\mu\nu} \simeq \exp\left(  \I\, G_{\mu\nu} a^2  
\right)
\eeq
Considering the expansion of the exponential on the right hand side,
the first non-trivial contribution to the imaginary part of its trace 
comes from the third order, which indeed is proportional to 
\beq
- \I\, {\rm Tr}\, G^3_{\mu\nu} = -i\, 
G^a_{\mu\nu}    G^b_{\mu\nu}    G^c_{\mu\nu}    
{\rm Tr} \left( T^a T^b T^c \right) = 
\frac{1}{4} (f^{abc} - \I\, d^{abc})\, 
G^a_{\mu\nu}    G^b_{\mu\nu}    G^c_{\mu\nu}   \; 
\eeq
where no sum over $\mu,\, \nu$ is understood.
A non-zero expectation value for the imaginary part of the plaquette,
therefore, corresponds to a non-zero 
charge odd, three gluon condensate, constructed in terms of the 
symmetric tensor $d^{abc}$:
\beq
\langle d^{abc}
G^a_{\mu\nu}    G^b_{\mu\nu}    G^c_{\mu\nu} \rangle \neq 0 \, .
\eeq

Such condensate,
which  is zero by charge conjugation symmetry in the normal QCD vacuum, 
will develop a non-zero value along some directions, e.g. for 
$(\mu, \nu) = (x,y)$ in case of a magnetic field in the $\hat z$ direction,
and can be considered as a non-trivial, charge-odd 
gluon background induced
by the external field. Of course the effect will be present only for
$N_c \geq 3$ colors, since for $SU(2)$ all traces are real and indeed
$d^{abc} = 0$.

The loop expansion of the determinant, examined above, is just a way
to show explicitly that such effects can be present, 
but cannot be quantitative, since the expansion is purely formal
(apart from the limit of large quark masses). 
A systematic investigation can and must be performed in the future
by means of lattice simulations. 

Similarly to 
charge conjugation symmetry,
one can investigate 
electromagnetic backgrounds which break explicitly 
the symmetry under CP, i.e. such that $\esb \neq 0$, 
and ask how CP violation propagates to the 
gluon sector, giving rise to an effective $\theta$ parameter.
Such phenomenon is in some sense complementary to 
the chiral magnetic effect and is related in general
to the effective 
QED-QCD interactions in the pseudoscalar 
channel~\cite{musak,elze1,elze2,mueller}.
It has been studied recently by a first exploratory 
lattice investigation~\cite{demane},
where numerical simulations at imaginary electric fields plus
analytic continuation have been exploited to determine
the susceptibility $\ceb$ 
of the QCD vacuum to CP-odd electromagnetic fields, defined by 
$\tef \simeq \ceb e^2 \esb$, obtaining
$\ceb \sim 7$ GeV$^{-4}$ for QCD with two staggered flavors and 
$m_\pi \sim 480$ MeV.

Finally, let us discuss a few other lines along which the effects of
electromagnetic background fields on the gluon sector
could be further investigated. The first consists in the 
determination of the effective action of given gluonic backgrounds,
using the techniques described in Section~\ref{subsec:3},
but in combination with non-trivial electromagnetic
external fields: that would give the possibility of studying if 
the latter can give rise to instabilities 
in the gluon sector, leading to the generation of non-trivial
gluonic backgrounds (as proposed e.g. in~\cite{galilo}).
The second regards the measurement of glueball masses and screening masses
in presence of magnetic backgrounds, which could give the possibility,
for instance, to test the proposal given in~\cite{anisotropic}
about the lowering of the confinement scale in presence
of a strong magnetic field. Finally, it would be important to 
perform a careful investigation about the
influence of magnetic backgrounds on the QCD equation of state,
as proposed in~\cite{ilgenfritz}.

\section{Conclusions}
\label{sec:conclusions}

Numerical simulations in presence of background external fields have been
considered since the early stages of Lattice QCD.
The last few years, however, have seen the development of considerable activity 
on the subject, which has been driven by the recent interest in theoretical
and phenomenological issues regarding the behavior of strongly interacting
matter in magnetic fields.

In this review, after a general overview about the formulation of 
lattice QCD in presence of magnetic or chromomagnetic background fields,
we have discussed the present status of lattice studies
regarding magnetic catalysis and the QCD phase diagram.
In the low temperature phase, consistent results have been
obtained, by different groups, which confirm magnetic catalysis
and the predictions coming from chiral perturbation theory
and some effective models. Good part of the enhancement 
of chiral symmetry breaking seems to be associated with modifications 
of gluon fields induced by the magnetic field via quark loops.
A chromagnetic background induces magnetic catalysis as well.

In the high temperature phase, consistent results have been obtained 
regarding the fact that a magnetic or chromomagnetic field shifts both chiral
symmetry restoration and deconfinement and that the two transitions
do not split, at least for fields up to $O(1)$~GeV$^2$.
A magnetic field also leads to a strengthening of the transition,
which however does not seem to turn into a strong first order,
at least for fields up to $O(1)$~GeV$^2$; it would be interesting however,
to further investigate the issue also in combination with other 
external parameters, like a baryon chemical potential, since that
could be relevant to the search for a critical endpoint in the QCD phase diagram.

Regarding the shift in the pseudocritical temperature induced by
a background magnetic field, existing studies have shown that
a modification of the lattice implementation of QCD can change 
the outcome, going from a slight increase 
(of the order a few \% for $|e| B$ of $O(1)$~GeV$^2$) 
to a decrease 
(of the order of 10~\% for $|e| B$ of $O(1)$~GeV$^2$) 
when an improved discretization and a physical quark mass spectrum
are used; in the latter case one also observes inverse magnetic catalysis,
i.e. a decrease of the chiral condensate as a function of $B$, in the 
high temperature, deconfined phase. A decrease of the pseudocritical temperature
is also observed in presence of a chromomagnetic background field.
A remaining question is which effect is more 
relevant to explain the discrepancy among different lattice studies,
i.e. whether lattice artifacts, or the unphysical quark spectrum, or both
are at its origin.

One of the aspects that emerges more clearly from present studies is the fact that
gluon fields are strongly modified by external electromagnetic fields, 
even though indirectly, by means of dynamical quark loops. We believe
that this aspect can and should be investigated more sistematically
by lattice simulations: in Sec.~\ref{gluodyn} we have discussed and suggested a few possible 
ways to do that by future studies.

\begin{acknowledgement}
The author is grateful to P.~Cea, L.~Cosmai, M.~Mariti, S.~Mukherjee, F.~Negro 
and F.~Sanfilippo for collaboration on some
of topics discussed in this review. He also acknowledges 
M.~Chernodub, G.~Endrodi, E.~Fraga, K.~Fukushima, V.~Miransky and M.~Ruggieri for
many useful discussions.
\end{acknowledgement}
%

\input{referenc}

\end{document}

%% file: referenc.tex
%
%
%